\documentclass[twocolumn, apj, numberedappendix, twocolappendix, iop]{openjournal}

\usepackage{natbib}
\usepackage{graphicx,amsmath,amssymb,amstext}
\usepackage{amsbsy,amsfonts,amsthm,color}
\usepackage[colorlinks,linkcolor=blue,citecolor=blue,urlcolor=blue ]{hyperref}
\usepackage[utf8]{inputenc}
\usepackage{float}

\usepackage{graphicx}
\usepackage{orcidlink}
\usepackage{booktabs}
\usepackage{xspace}

\usepackage{upgreek}

\newcounter{FootnoteA}

\begin{document}

    \title{The Cosmological Simulation Code \textsc{OpenGadget3} -- Implementation of Self-Interacting Dark Matter\vspace{-4ex}}
    \shorttitle{\textsc{OpenGadget3} -- SIDM Implementation}
    \author{Moritz S.\ Fischer$^{1,2,*}$\orcidlink{0000-0002-6619-4480}}
    \author{Marc Wiertel$^{3}$\orcidlink{0009-0000-0328-9901}}
    \author{Cenanda Arido$^{4,5}$\orcidlink{0009-0002-5399-4742}}        
    \author{Yashraj Patil$^{6,7}$\orcidlink{0009-0005-0275-4353}}
    \author{Antonio Ragagnin$^{8}$\orcidlink{0000-0002-8106-2742}}
    \author{Klaus Dolag$^{2,9}$\orcidlink{0000-0003-1750-286X}}
    \author{Marcus Brüggen$^{10}$\orcidlink{0000-0002-3369-7735}}
    \author{Mathias Garny$^{5}$\orcidlink{0000-0003-4056-6802}}
    \author{Andrew Robertson$^{11}$\orcidlink{0000-0002-0086-0524}}
    \author{\mbox{Kai Schmidt-Hoberg$^{12,13}$}\orcidlink{0000-0002-9851-9091}}
    \shortauthors{M.\ S.\ Fischer et al.}
    \email{$^*$moritz.fischer@dipc.org}
    \affiliation{$^1$Donostia International Physics Center (DIPC), Paseo Manuel de Lardizabal 4, 20018 Donostia-San Sebastián, Spain}
    \affiliation{$^2$Universitäts-Sternwarte, Fakultät für Physik, Ludwig-Maximilians-Universität München, Scheinerstr.\ 1, D-81679 München, Germany}
    \affiliation{$^3$Institute for Theoretical Particle Physics (TTP), Karlsruhe Institute of Technology (KIT), D-76128 Karlsruhe, Germany}
    \affiliation{$^4$The Oskar Klein Centre, Department of Physics, Stockholm University, Albanova University Center, SE-106 91 Stockholm, Sweden}
    \affiliation{$^5$Physik Department T31, Technische Universit\"at M\"unchen, James-Franck-Straße 1, D-85748 Garching, Germany}
    \affiliation{$^6$College of Science and Engineering, University of Minnesota, 117 Pleasant St, Minneapolis, MN 55455, USA}
    \affiliation{$^7$Birla Institute of Technology and Science-Pilani, K. K. Birla Goa campus, NH-17B, Zuarinagar, Goa 403726, India}
    \affiliation{$^8$INAF - Osservatorio Astronomico di Trieste, via G. Tiepolo 11, 34143 Trieste, Italy}
    \affiliation{$^{9}$Max-Planck-Institut f\"ur Astrophysik, Karl-Schwarzschild-Str.~1, D-85748 Garching, Germany}
    \affiliation{$^{10}$Hamburger Sternwarte, Universität Hamburg, Gojenbergsweg 112, D-21029 Hamburg, Germany}
    \affiliation{$^{11}$Carnegie Observatories, 813 Santa Barbara Street, Pasadena, CA 91101, USA}
    \affiliation{$^{12}$Deutsches Elektronen-Synchrotron DESY, Notkestr.~85, 22607 Hamburg, Germany}
    \affiliation{$^{13}$CP3-Origins, University of Southern Denmark, Campusvej 55, DK-5230 Odense M, Denmark}

    \begin{abstract}
        Dark matter (DM) could be subject to non-gravitational self-interactions which is relevant to resolve potential problems of cold DM on small scales. Their impact on astrophysical objects such as galaxies and galaxy clusters allows for constraining the strength of this scattering and eventually further properties of the cross-section. To model self-interacting dark matter (SIDM), $N$-body simulations are a crucial tool widely employed by the SIDM community.
        In this paper, we describe the SIDM implementation in the cosmological hydrodynamical $N$-body code \textsc{OpenGadget3} and release it to the public. It is capable of simulating elastic scattering for various differential cross-sections, including strongly anisotropic cross-sections. Beyond single-species models, the code also allows simulating a two-species model with cross-species interactions. In addition to describing the numerical schemes for modelling various flavours of SIDM, we discuss the technical challenges of implementing them. Moreover, we demonstrate through several test problems that \textsc{OpenGadget3} can accurately simulate DM self-interactions.
        Furthermore, we assess the performance of the code and provide scaling tests. Lastly, we highlight remaining challenges in the context of SIDM and describe directions for improving the current state of the art.
    \end{abstract}
  

    \maketitle

\section{Introduction} \label{sec:introduction}
Self-interacting dark matter (SIDM) is a class of particle physics models that provides candidates to explain the nature of dark matter (DM). In contrast to the collisionless cold dark matter (CDM) of the cosmological standard model, $\Lambda$CDM, SIDM involves interactions beyond gravity. DM particles can scatter off each other, which potentially alters the DM distribution on small scales, i.e.\ galactic scales. At the same time, the large-scale structure remains unchanged compared to that of CDM. Historically, SIDM had been introduced by \cite{Spergel_2000} with the motivation of providing a mechanism to reduce the abundance of satellite galaxies and reduce the inner density of DM halos.\footnote{See also \cite{Carlson_1992} for even earlier work in a more cosmological context.} But SIDM has been studied in various contexts, and a variety of particle physics models have been explored. For reviews on SIDM, we refer the reader to \cite{Tulin_2018} and \cite{Adhikari_2025b}.

Various observational probes have been employed to constrain the strength of DM self-interactions. This includes the size of DM density cores across a wide range of halo masses \citep[e.g.][]{Sagunski_2021, Yang_2024S, Diego_2026}, the shapes of DM halos \citep[e.g.][]{Peter_2013}, and an offset between galaxies and DM in merging galaxy clusters \citep[e.g.][]{Harvey_2015, Wittman_2023}.
Deriving all these constraints made it necessary to model the effect of DM scattering. In order to derive constraints, often semi-analytical models are employed, as they are computationally cheap. At the same time, $N$-body simulations have been proven to be crucial for understanding the astrophysical phenomenology of SIDM from first principles and for validating and calibrating semi-analytic models.

Collisionless $N$-body simulations are a widespread tool in astrophysics to study, for example, the formation of the large-scale structure or the evolution of galaxies and galaxy clusters.
The matter distribution is represented by means of $N$ particles, where each represents a patch of phase-space. Hence, it is typically assumed that a numerical particle represents numerous physical particles that are close in phase-space. The numerical particles are characterised by their mass, position, and velocity. Those quantities are evolved over time by discretising and solving the differential equations governing the physical system. In practice, this includes gravity, which allows one to compute the acceleration that each particle experiences. Numerical schemes for $N$-body simulations of self-gravitating systems have been developed for several decades \citep[for a review see, e.g.][]{Dehnen_2011}.

In contrast, SIDM $N$-body simulations are younger. The first one was run by \cite{Burkert_2000}, where the modelling of the DM scattering is based on a stochastic process. A probability for a particle to interact with its nearest neighbour is computed. In detail, it depends on the local density and the particle's velocity in the rest frame of the halo. The Monte-Carlo approach by \cite{Burkert_2000} draws a random number and compares it to the probability to decide whether the particle interacts with the neighbouring particle.
An important improvement at that time was made by \cite{Kochanek_2000}; they used the actual phase-space distribution for the interactions, i.e., computed the relative velocity of the numerical particles and used it in computing the interaction probability.
More similar schemes were quickly implemented \citep[][]{Yoshida_2000b, Dave_2001} and some already included a velocity-dependent cross-section \citep{Colin_2002, D_Onghia_2003}. 
In addition to employing a Monte-Carlo scheme, very strong cross-sections have also been modelled using smoothed particle hydrodynamics (SPH) \citep[][]{Moore_2000, Yoshida_2000a}.

Further SIDM implementations followed in various astrophysical $N$-body codes \citep[e.g.][]{Koda_2011, Vogelsberger_2012, Rocha_2013, Fry_2015, Robertson_2017a, Robles_2017, Vogelsberger_2019, Banerjee_2020, Yang_2020, Correa_2022, Fischer_2021a, Ebisu_2022, Valdarnini_2024, Bosch_2026}.
Although almost all SIDM $N$-body codes are based on a Monte-Carlo scheme, there is one notable exception by \cite{Huo_2020}.

Many particle physics models are described by a differential cross-section that is velocity and angular dependent, and the two dependences are often not separable. The first implementation that allows one to model such a differential cross-section was written by \cite{Robertson_2017b} and applied to simulating merging galaxy clusters. Later, other implementations of full differential cross-sections followed, e.g.\ by \cite{Banerjee_2020, Correa_2022, Yang_2022D}, or the one in \textsc{OpenGadget3} that we describe here.

A common problem in SIDM $N$-body simulations occurs when a particle undergoes multiple scatterings per time step without updating the particle's velocity between the interactions. The exchange of kinetic energy between the numerical particles cannot be treated in a simple cumulative manner, as doing so leads to an error in energy conservation, introducing an artificial heating term \citep{Robertson_2017a}.
Fully resolving this issue is difficult, as one needs to ensure that the velocity of a particle is updated after an interaction before undergoing another scattering event within the same time step. In particular, for parallel computations, this can require a fairly sophisticated parallelisation scheme. \cite{Robertson_2017a} have largely improved on this issue \citep[see also][]{Robertson_2017T}. However, it has now been fully resolved by \cite{Fischer_2021a}, with \cite{Valdarnini_2024} presenting an alternative approach that fulfils the same purpose. 

Another challenge that had persisted for a while was the simulation of strongly forward-dominated cross-sections.\footnote{Small-angle scattering needs to be much more frequent than large-angle scattering to have a comparable effect on the DM distribution. As a consequence, strongly forward-dominated cross-sections are also referred to as frequent SIDM (fSIDM) in contrast to rare SIDM (rSIDM).}
Within the usual SIDM schemes, they require extremely small time steps in order to keep the interaction probability well below unity, effectively leading to prohibitively high computational costs.
An approach to avoid this problem is to simply exclude those small-angle scattering events, as, for example, done by \cite{Robertson_2017b}.
However, those events could be so frequent that they contribute significantly to the momentum or energy transfer between the particles.
An attempt to study such a cross-section in the context of a merging galaxy cluster was made by \cite{Kahlhoefer_2014}. They had built a semi-analytic approach by combining an $N$-body code with an effective drag-like description for small-angle scattering. 
Another way of approaching the problem was undertaken by \cite{Kummer_2019}. They aimed to model the evolution of an isolated halo by effectively describing small-angle scattering as heat conduction within an $N$-body simulation. However, the problem was only fully resolved by \cite{Fischer_2021a}, who presented the first scheme that allowed one to model such a cross-section from first principles within $N$-body simulations.
It applies an effective drag force and perpendicular momentum diffusion to pairs of neighbouring $N$-body particles, building on analogous methods known from seminal works by \cite{Einstein_1905, Smoluchowski_1906} related to Brownian motion.
The drag force acting along the axis defined by the relative velocity leads to a deceleration that is applied deterministically in each time step, representing the cumulative effect of a large number of small-angle scatterings. The momentum diffusion within the plane perpendicular to this axis is described as a stochastic process, specifically regarding the direction along which a perpendicular momentum kick is applied, while its magnitude is fixed by energy conservation. Overall, this leads to a deterministic effective scattering angle that depends on the cross-section, relative velocity and density in accordance with the microscopic dynamics.
 
Recently, \cite{Arido_2025} combined the schemes for large- and small-angle scattering. This allowed -- for the first time -- the efficient simulation of differential cross-sections that fall in both regimes. The underlying idea is to introduce a critical angle; the parts of the differential cross-section with scattering below this angle are treated with the small-angle scattering scheme, and the other ones with the larger-angle scattering scheme. This extends the range of models that can be practically simulated.

Various SIDM models have been studied so far with $N$-body simulations. This does not only concern the exploration of various angular and velocity dependencies, such as for models of resonant scattering \citep{Tran_2024}. But it goes beyond the assumption of elastic scattering or having only a single DM particle species.
A few simulations exist with models where the scattering is no longer elastic, but a fraction of the kinetic energy is lost per interaction. Such models have been studied by
\cite{Huo_2020, Shen_2021, Shen_2024, Xiao_2021}. Closely related are models with multiple states, where energy can be stored in excited states. While particles interact, they may change between the ground and excited state. \cite{Vogelsberger_2019, Chua_2020, ONeil_2023, Leonard_2024, Low_2026} have investigated such models.
Recently, there has also been interest in models with two species with interactions between the species \citep{Patil_2025, Yang_2025a, Yang_2025b}.
These variations on the classic elastic SIDM model give rise to diverse phenomenology, with many interesting effects that can potentially be probed observationally. 

A fascinating consequence that many SIDM models have in common is the gravothermal collapse of DM halos. The self-interactions may cause the halo to lose energy, by transporting it to larger radii. As a consequence, the central DM density grows, and -- given the negative heat capacity of gravitationally self-bound objects -- the centre also heats up, accelerating the gravothermal evolution. Motivated by observations that may provide hints for surprisingly dense objects \citep[e.g.][]{Vegetti_2010, Meneghetti_2020, Minor_2021, Granata_2022, Dutra_2025, Enzi_2025, Li_2025, Minor_2025, Tajalli_2025, Kollmann_2026, Vegetti_2026}, these later stages of gravothermal evolution have received a lot of attention recently~\citep[e.g.][]{Yang_2021, Nadler_2023, Gad-Nasr_2024, Yu_2026, Zhang_2025, Engelhardt_2026, Fischer_2026, Gu_2026, Kamionkowski_2026, Kong_2026, Li_2026, Mace_2026}. However, it has transpired that these stages are difficult to model with $N$-body codes \citep[e.g.][]{Yang_2022D, Zhong_2023, Mace_2024, Palubski_2024}. Not all of these problems are directly related to the modelling of SIDM, but some are also present in simulations of collisionless DM, but become more severe for the dense halos produced by the gravothermal evolution \citep{Fischer_2024b}. Despite some concerns, no fundamental problems were encountered in modelling SIDM with $N$-body simulations. Although modelling the late gravothermal evolution accurately is very expensive and requires increasing resolution and smaller time steps, the deeper one simulates into the gravothermal collapse \citep{Fischer_2025b}.

The underlying problem of modelling SIDM, i.e.\ solving the Boltzmann equation, had been addressed well before any SIDM $N$-body simulation had been run. For problems rather different from SIDM, generally referring to rarefied gases, for example, in the context of the reentry of a spaceship, Direct Simulation Monte-Carlo (DSMC) schemes have been developed \citep{Bird_1963, Bird_1994, Oran_1998, Pareschi_2001, Gimelshein_2019}. They commonly rely on grid cells instead of performing a neighbour search and employing kernel functions, as is done in state-of-the-art SIDM $N$-body simulations.
As a consequence, adaptive meshing techniques have been developed for DSMC \citep[e.g.][]{Kim_2004, Wu_2004}.
However, there have also been studies that developed meshless DSMC methods \citep[e.g.][]{Olson_2008, Shen_2023}, being more closely related to the schemes being usually explored for SIDM.
A first study with a grid-based DSMC method in the context of SIDM was undertaken by \cite{Gurian_2025}. They studied the gravothermal evolution of an isolated spherically symmetric SIDM halo and took advantage of the symmetries of the set-up by reducing the dimension of the DSMC grid from three to one. A closely related work with a similar scheme followed by \cite{Kamionkowski_2026}.

The aim of this paper is to provide an overview of the SIDM implementation in the cosmological hydrodynamical $N$-body code \textsc{OpenGadget3} \citep{Dolag_2026}. This was introduced by \cite{Fischer_2021a} and refined in several later studies. Initially, it only enabled simulating elastic velocity-independent scattering for isotropic or extremely forward-dominated cross-sections. With \cite{Fischer_2022}, comoving integration was added, to enable cosmological simulations. Moreover, the implementation was extended to simulate velocity-dependent interactions and got a more efficient message passing interface (MPI) parallelisation \citep{Fischer_2024a}. 
\cite{Arido_2025} enhanced the capabilities of simulating anisotropic cross-sections. In particular, it became possible to simulate a differential cross-section that is forward-dominated but also has non-negligible larger-angle, a situation that is typically encountered for scatterings described by Coulomb or Yukawa potentials.
Another recent extension is the introduction of a two-species model with interactions between the two species \citep{Patil_2025}.
In addition to the publications that have improved or extended the SIDM implementation in \textsc{OpenGadget3}, the module has also been used in several studies \citep{Fischer_2021b, 
Fischer_2023b, Fischer_2024c, Ragagnin_2024, Sabarish_2024, Sabarish_2025}.
In particular, \citep{Fischer_2024b} and \citep{Fischer_2025b} have investigated the numerical properties of \textsc{OpenGadget3} in the context of the gravothermal collapse of SIDM halos and demonstrated its accuracy.

Here, we will go beyond what has been presented in the above-mentioned publications and release\footnote{The full \textsc{OpenGadget3} code was made public with the main release paper \cite{Dolag_2026} and is available at \url{https://gitlab.lrz.de/AstroCodes/OpenGadget3}} a version that is additionally capable of simulating full differential cross-sections. This, in particular, includes models where the velocity and angular dependence are not separable. Moreover, we describe for the first time an Open Multi-Processing (OpenMP) parallelisation for the SIDM computations, which is part of the release. Beyond the code itself, we show several test problems, some of them for the first time, that can be used to test SIDM simulation codes. Importantly, we investigate how the SIDM implementation scales with the number of processors used and provide advice on choosing the simulation parameters.

The remainder of the paper is structured as follows. In Sect.~\ref{sec:numerical_method} we describe the numerical method on which the SIDM implementation of \textsc{OpenGadget3} is based. Here, we also discuss the implementation and related challenges, in particular when it comes to the parallelisation of the computations. It is followed by a series of test problems in Sect.~\ref{sec:test_problems}, where we demonstrate that we can produce accurate results for problems with a known solution.
In addition, we also simulate more astrophysically relevant set-ups to test the scaling of the SIDM module as well as comment on the particular challenge of simulating the collapse phase of the gravothermal evolution. In Sect.~\ref{sec:discussion}, we discuss shortcomings of the current implementation and avenues for improving upon the current state of the art. Finally, in Sect.~\ref{sec:conclusion}, we summarise and conclude.
Additional information is provided in the Appendix.

\section{Numerical Method} \label{sec:numerical_method}

The systems we aim to describe consist of an enormous number of dark matter (DM) particles. In this regime, a statistical description becomes appropriate: instead of tracking individual particles, one may consider the smooth phase-space density $f(\mathbf{x}, \mathbf{v}, t)$, which is well defined only in the limit of many particles. In this continuum limit, the evolution of the system is deterministic and governed by the Boltzmann equation with self-gravity.

The underlying goal of the numerical method presented in this section is to model the impact of DM self-interactions on astrophysical objects. Since their dynamics are primarily driven by gravity, we seek to solve the Vlasov--Poisson equation supplemented by a collision term, i.e.\ the Boltzmann equation including self-gravity,
\begin{equation} \label{eq:vlasov_poisson}
    \frac{\partial f}{\partial t}
    + \mathbf{v} \cdot \nabla_x f
    - \nabla_x \Phi \cdot \nabla_v f
    = \left(\frac{\partial f}{\partial t}\right)_\mathrm{coll} \;.
\end{equation}
Here, $f$ denotes the phase-space density, $\mathbf{v}$ the velocity, and $\Phi$ the gravitational potential. The self-interactions are encoded in the collision term on the right-hand side, which depends on the differential cross-section specified by the underlying particle physics model.

Although microscopic DM scatterings are stochastic processes, the macroscopic systems of interest behave deterministically in the limit of large particle numbers. Here, Eq.~\eqref{eq:vlasov_poisson} provides a valid statistical description.

Complications arise, however, when employing an $N$-body method to model SIDM. The discretization of the phase-space density into a finite number of simulation particles (see Sect.~\ref{sec:numerical_representation}) reintroduces stochasticity at the numerical level. As a consequence, the collision term can no longer be implemented in a purely deterministic manner. Instead, the numerical scheme relies on a Monte-Carlo procedure whose stochastic realization converges, in the large-number limit, to the deterministic evolution of the phase-space density described by Eq.~\eqref{eq:vlasov_poisson}.

\subsection{Numerical representation and kernel overlap} \label{sec:numerical_representation}

The physical entities, in our case, the DM particles, are represented by $N$ numerical particles. They represent all the mass, momentum, and energy of the DM. Each numerical particle is characterised by a mass, $m_\mathrm{n}$,\footnote{Here, we use the label ``n'' to distinguish the numerical particle mass from the mass of the physical particles.} a position, $\mathbf{x}$, and a velocity, $\mathbf{v}$. One numerical particle can be interpreted as representing a phase-space patch of physical DM particles. In this picture, all the DM particles of the phase-space patch have the same velocity, as each numerical particle has only a single velocity.

\begin{figure}
    \centering
    \includegraphics[width=0.75\columnwidth]{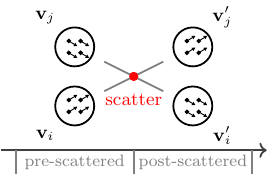}
    \caption{Illustration of the numerical representation and the scattering process for the SIDM scheme. The black circles denote numerical particles, with the black dots inside illustrating that they consist of numerous physical DM particles. The numerical particles evolve from the pre-scattered state (left) to the post-scattered state (right) by changing the direction of their momentum vectors in their centre-of-mass frame.}
    \label{fig:sketch}
\end{figure}

For SIDM, every numerical particle is additionally assigned a kernel.
It is introduced to obtain a local density estimate used for the modelling of the self-interactions. Although the kernel function does not represent a fundamental property of the physical system, it can be useful for the derivation of the SIDM scheme to interpret it as a description of the spatial extent of the represented DM, i.e.\ the phase-space patch's distribution in configuration space. The kernel has a positive finite size $h$, implying that the kernel function $W(d,h)$ becomes zero at distances $d \geq h$. The density represented by a single numerical particle can be understood as $\rho = W(d,h) \, m_\mathrm{n}$. We determine $h$ for each particle adaptively by searching for the $N_\mathrm{ngb}$ nearest neighbours, i.e.\ $h$ is a radius that includes the nearest $N_\mathrm{ngb}$ particles.
We illustrate the numerical representation and the interaction of two numerical particles in Fig.~\ref{fig:sketch}.

The effect of self-interactions is computed based on pairs of neighbouring particles. Pairs that are close enough can interact; the criterion is $|\mathbf{x}_i - \mathbf{x}_j| < h_i + h_j$.
To compute the interaction between a pair of particles, we employ a geometric factor $\Lambda_{ij}$. It is given by the volume integral over the product of the kernels of the two particles.
\begin{equation} \label{eq:kernel_overlap}
    \Lambda_{ij} = \int W(|\mathbf{x}-\mathbf{x}_i|, h_i) \, W(|\mathbf{x}-\mathbf{x}_j|, h_j) \, \mathrm{d}^3\mathbf{x} \,.
\end{equation}
This kernel overlap is important for determining how strong the interaction between the two particles is, as we describe in detail in Sect.~\ref{sec:rare_scatter} and~\ref{sec:freq_scatter}.

In principle, various functions can be used for $W(d,h)$. They have to be: a) non-negative everywhere, $W(d, h) \geq 0 \, \forall \, d \in \mathbb{R}^+_0$; b) zero beyond $h$, $W(d, h) = 0 \, \forall d \geq h$; c) normalised, i.e.\ $1 = \int W(\mathbf{x}, h)\,\mathrm{d}^3\mathbf{x}$, implying that they are integrable. However, the kernel function does not need to be differentiable, as is required for SPH, and it does not even need to be continuous.

In earlier tests, we tried various kernel functions but found hardly any difference between them \citep{Fischer_2021a}.
In practice, we employ the cubic spline kernel introduced by \cite{Monaghan_1985}, which is also a popular choice for SPH simulations.
For our studies, we use a scaled version,
\begin{equation}
    W(d,h)= \left\lbrace \begin{matrix} \frac{8}{\uppi \, h^3} \left[1-6 \, q^2 \, (1- q)\right]&\, \textnormal{if}&0\leq q < 0.5,\\ \frac{16}{\uppi \, h^3}[1-q]^3&\, \textnormal{if}&0.5\leq q < 1, \\ 0 &\, \textnormal{if}&1 \leq q . \end{matrix} \right.
\end{equation}
with $q:=d/h$.
Nevertheless, various other kernel functions are implemented in \textsc{OpenGadget3} and can easily be used instead of the cubic spline kernel.
These include a top-hat kernel, the quintic spline kernel \citep{Morris_1996}, the Wendland $C^4$ and $C^6$ kernels \citep{Wendland_1995}.

In \textsc{OpenGadget3}, the kernel overlap (Eq.~\eqref{eq:kernel_overlap}) is computed numerically and for each pair interpolated from a lookup table. This makes it easy to implement new kernel functions. For more information on the kernel overlap computations, we refer the reader to appendix~A by \cite{Fischer_2021a}.
A few other SIDM implementations are also based on the kernel overlap integral \citep{Rocha_2013, Correa_2022}.

We note that the formulation of the geometric factor given by Eq.~\eqref{eq:kernel_overlap} is a fully symmetric formulation. More generally speaking, we follow the idea that the numerical difference equations should be derived in a way such that they reflect the symmetries of the underlying differential equations, and no ad-hoc symmetrisation is needed.

In several other implementations, an asymmetric formulation is used. Often, the kernel function is simply evaluated at the position of the other particle \citep[see sect.~II.~B. by][]{Adhikari_2025b}. Here, the geometric factor can take different values for the two particles of a pair due to different kernel sizes. To deal with this, these schemes de facto symmetrise the geometric factor to obtain a new value used for the pair. This is based on the assumption that the deviation of the geometric factor of each particle from the symmetrised value is random and thus cancels out over many interactions. However, if it were no longer random but correlated with properties of the numerical particles, this would lead to a systematic under- or overestimate of the geometric factor and thus would render the numerical scheme inconsistent with the physical scattering of DM particles. We avoid this potential issue right from the beginning by using a fully symmetric formulation.

\subsection{Angle-averaged cross-sections}
Different angle-averaged cross-sections play a central role in the study of SIDM scatterings and in the construction of numerical schemes. Depending on the physical process under consideration, different angular averages emphasise distinct physical effects. In the following, we introduce the specific averages relevant for the SIDM implementation in \textsc{OpenGadget3}.

We begin with the total cross-section, which is proportional to the rate of scattering events,
\begin{equation} \label{eq:total_cross_section}
\sigma=\int_{-1}^{1} \frac{\mathrm{d} \sigma}{\mathrm{d} \cos \theta_{\mathrm{cms}}} \, \mathrm{d} \cos \theta_{\mathrm{cms}} \, .
\end{equation}
The differential cross-section in the centre-of-mass system (cms) is given by $\mathrm{d}\sigma / \mathrm{d}\cos\theta_\mathrm{cms}$ and $\theta_\mathrm{cms}$ is the scattering angle. We note that the factor of $2\uppi$ from the integration of the azimuth angle $\varphi_\mathrm{cms}$ cancels out, assuming the cross-section to be independent of $\varphi_\mathrm{cms}$, as is typically the case in DM models.

Often, however, one is not primarily interested in the total scattering rate, but rather in other physical questions, such as how efficiently energy or momentum is transferred between different systems, or how rapidly thermalisation occurs. These questions can often be addressed using an effective cross-section and do not strongly depend on the details of the underlying differential cross-section. One example is the viscosity cross-section, which weights scattering events by their efficiency in redistributing transverse momentum relevant e.g.\ for thermal conduction.
\begin{equation} \label{eq:viscosity_cross_section}
\sigma_\mathrm{V} = \int_{-1}^{1} \frac{\mathrm{d} \sigma}{\mathrm{d} \cos \theta_{\mathrm{cms}}}\sin^2\theta_{\mathrm{cms}} \, \mathrm{d} \cos \theta_{\mathrm{cms}} \, .
\end{equation}
It has been used to compare the strength of differential cross-sections with various angular dependences. For example, it allows to quantify the strength of the self-interaction for the gravothermal evolution of an isolated halo very well \citep[e.g.][]{Yang_2022D, Fischer_2025b}.

Another relevant quantity is the modified momentum transfer cross-section \citep{Kahlhoefer_2017, Robertson_2017b},
\begin{equation} \label{eq:mod_momentum_transfer_cross_section}
\sigma_\mathrm{\Tilde{T}}= \int_{-1}^{1} \frac{\mathrm{d} \sigma}{\mathrm{d} \cos \theta_{\mathrm{cms}}}\left(1-|\cos \theta_{\mathrm{cms}}| \right) \mathrm{d} \cos \theta_{\mathrm{cms}} \, .
\end{equation}
Here, the effect of forward ($\theta_{\mathrm{cms}} \approx 0$) and backward ($\theta_{\mathrm{cms}} \approx \uppi$) scattering is down-weighted, similar to the case of the viscosity cross-section.
It describes the momentum transfer between different systems appropriate for identical particle scattering. In particular this form takes into account that a scattering angle $\theta$ is physically identical to $\uppi - \theta$, i.e.\ corresponds to the same physical final state, just with the particles relabelled. As an example a back-scattering event would not change the physical state for indistinguishable particles.\footnote{In the context of this paper, the distinguishability of particles refers to their physical behaviour relevant for the evolution of the simulated systems. Or in other words, particles are distinguishable if exchanging the species would lead to a different evolution of the physical system. This can be because the species vary in their physical DM particle mass, or differ in other interactions, for example, with themselves or baryons.} This form is also appropriate for particle anti-particle scatterings, if one is interested in the effective momentum transfer.

For forward-dominated scattering, the total cross-section is typically much larger than the momentum-transfer or viscosity cross-sections, reflecting the large probability of small-angle scatterings. A well-known example is Coulomb scattering, for which the total cross-section is formally infinite, while $\sigma_\mathrm{\Tilde{T}}$ and $\sigma_\mathrm{V}$ are finite. The physical impact of self-scattering is in this case captured by Eqs.~\eqref{eq:viscosity_cross_section} and~\eqref{eq:mod_momentum_transfer_cross_section}. For (nearly) isotropic scattering, on the other hand, all three cross-sections are generically of comparable size, up to order-one factors. In Table~\ref{tab:cross-sections} we summarise the conversion factors between different angle-averaged cross-sections for two different angular dependencies. Namely, isotropic scattering and the limit of an extremely forward-dominated cross-section. In terms of a differential cross-section, this limit can, for example, be expressed as
\begin{align} \label{eq:cross-section_fwd_indist}
    \left.\frac{\mathrm{d}\sigma}{\mathrm{d}\cos \theta_{\mathrm{cms}}}\right|_\mathrm{fwd}= \lim_{\epsilon\to 0} \,&\frac{\sigma_\mathrm{\tilde{T}}}{2 \,\ln(\epsilon^{-2})} \nonumber\\
    &\frac{(3\cos^2\theta_\mathrm{cms}+1) \, \epsilon^{-4} + 4 \, \epsilon^{-2} +4}{\left((1-\cos^2\theta_\mathrm{cms})\,\epsilon^{-4} + 4 \, \epsilon^{-2} + 4\right)^2} \,,
\end{align}
assuming indistinguishable particles. This expression is motivated by the Born cross-section for a Yukawa potential, with the limit $\epsilon\to 0$ corresponding to the case of a Coulomb potential, i.e.\ a massless force mediator. It can also be viewed as approximating a sharply peaked differential cross-section that approaches a sum of Dirac-distributions located at $\cos\theta_\mathrm{cms}=\pm 1$, with momentum-transfer cross-section Eq.~\eqref{eq:mod_momentum_transfer_cross_section} held fixed as $\epsilon\to 0$, while the total cross-section tends to infinity. This type of self-scattering can be efficiently simulated by the code developed in \citep{Fischer_2021a} and presented in this work based on the effective drag-force description.

In the literature, various normalization conventions exist, which differ from Eqs.~\eqref{eq:viscosity_cross_section} and~\eqref{eq:mod_momentum_transfer_cross_section} by a constant prefactor. While we use the convention from above, we introduce for convenience a notation for another commonly used normalization, related to ours as  $\sigma_\mathrm{V,norm} \equiv 3/2 \, \sigma_\mathrm{V}$ and $\sigma_\mathrm{\Tilde{T}, norm} \equiv 2 \, \sigma_\mathrm{\Tilde{T}}$. The rationale of considering this alternative convention is that both $\sigma_\mathrm{V,norm}$ and $\sigma_\mathrm{\Tilde{T}, norm}$ coincide with the total cross-section for isotropic scattering.

\begin{table}
    \centering
    \begin{tabular}{c|c|c}
        $c \equiv \sigma_X / \sigma_\mathrm{V,norm}$ & fwd.\ & iso.\ \\ \hline 
        $\sigma$ & $\infty$ & $1$ \\
        $\sigma_\mathrm{\Tilde{T}}$ & $1/3$ & $1/2$ \\
        $\sigma_\mathrm{\Tilde{T},norm}$ & $2/3$ & $1$ \\
        $\sigma_\mathrm{V}$ & $2/3$ & $2/3$ \\
        $\sigma_\mathrm{V,norm}$ & $1$ & $1$
    \end{tabular}
    \caption{Conversion between angle-averaged cross-sections. Relative size of various angle-averaged cross-sections for the case of forward-dominated scattering as well as for an isotropic differential cross-section, respectively.
    The left column states the angle-averaged cross-sections for which the relative size factors are given in the middle column in the case of a forward-dominated cross-section, and in the right column for isotropic scattering.
    }
    \label{tab:cross-sections}
\end{table}

While the modified momentum transfer cross-section given by Eq.~\eqref{eq:mod_momentum_transfer_cross_section} is the preferred choice for models with indistinguishable particles (including particle anti-particle scatterings), we employ the momentum transfer cross-section for distinguishable particles\footnote{For particle physics models for which the fSIDM description (see Sect.~\ref{sec:freq_scatter}) is applicable, the differential cross-section is strongly enhanced for either $(i)$ both $\theta\to 0$ and $\theta\to \uppi$ (for scattering of identical particle species) or $(ii)$ only for $\theta\to 0$ (for distinguishable particles). In case $(i)$ $\sigma_\mathrm{\Tilde{T}}$ is the relevant cross-section for the fSIDM description, while in case $(ii)$ both $\sigma_\mathrm{\Tilde{T}}$ and $\sigma_\mathrm{T}$ could be used. Indeed, for forward-dominated scattering of distinguishable particles the difference between these two angle-averaged cross-sections is small by assumption, with their difference being of the same order as the small contribution from large-angle scatterings that are assumed to be negligible when using the pure fSIDM description. When adopting the hybrid scheme from~\cite{Arido_2025}, the same applies to those scatterings treated via the fSIDM approach.},
\begin{equation} \label{eq:momentum_transfer_cross_section}
\sigma_\mathrm{T}= \int_{-1}^{1} \frac{\mathrm{d} \sigma}{\mathrm{d} \cos \theta_{\mathrm{cms}}}\left(1-\cos \theta_{\mathrm{cms}} \right) \mathrm{d} \cos \theta_{\mathrm{cms}} \, .
\end{equation}
We note, that differences compared to the modified momentum transfer cross-section (Eq.~\eqref{eq:mod_momentum_transfer_cross_section}) arise only from scattering angles $\theta_\mathrm{cms} > \uppi / 2$.
The momentum transfer cross-section is, for example, relevant in a model with two DM species where it is also the better-suited quantity to measure the strength of the interactions across the species compared to the viscosity cross-section, which is a common measure for interactions within a species \citep[e.g.][]{Patil_2025}.
We note that one can also express the forward-dominated limit for distinguishable particles as a differential cross-section, for example,
\begin{equation}  \label{eq:cross-section_fwd_dist}
    \left.\frac{\mathrm{d}\sigma}{\mathrm{d}\cos \theta_{\mathrm{cms}}}\right|_\mathrm{fwd}=\lim_{\epsilon\to 0} \frac{\sigma_\mathrm{T}}{4 \, \ln(\epsilon^{-2})} \frac{1}{\left(\epsilon^2 + \sin^2{(\theta_\mathrm{cms}/2)}\right)^2} \,.
\end{equation}

In the following two sections we consider the two algorithms provided by the code, adapted to differential cross-sections dominated by large- and small-angle scatterings, respectively. The former leads to a large momentum transfer per scattering, and is typically characterized by a relatively low total scattering rate, while the latter efficiently describes frequent small-angle scatterings with a large total scattering rate but small momentum transfer per scattering.

\subsection{Rare self-interactions} \label{sec:rare_scatter}

In the following, we describe the modelling of rare self-interactions. By this we refer to a regime in which the DM scattering rate is low enough such that one can assume that a physical particle represented by a numerical particle will not scatter more than once per pair-wise interaction of numerical particles.\footnote{In this context, we want to highlight the discussion given under point four in sect.~4 by \cite{Fischer_2024b} related to this assumption.} But it could still scatter multiple times per time step as multiple numerical pairs are considered. We note that most SIDM implementations require a stricter criterion, where a particle is allowed to interact only once per time step. But thanks to its parallelisation, the implementation in \textsc{OpenGadget3} allows relaxing this condition; it must hold only on the basis of a single pair and not all pairs.
This allows us to use larger time steps in situations where the time step size is limited by the self-interactions, e.g.\ in the central region of gravothermally collapsing SIDM halos.

In the picture of a numerical particle representing a phase-space patch, physical particles would scatter in all directions when two phase-space patches interact. The physical particles represented by numerical particles are located next to each other in phase-space before the interaction. After the interaction, they could be distributed over a substantial part of the phase-space. Some of them remain unchanged, i.e.\ the ones that do not scatter. Unfortunately, this post-interaction distribution of physical particles can no longer be represented by a single numerical particle, as it represents only a single velocity. As a consequence, we employ a Monte-Carlo approach by picking only one of the possible post-interaction locations in phase-space. Each of these locations is associated with a probability density that we should satisfy when selecting the post-interaction location. This is done in two steps: Firstly, we decide if the phase-space location changes at all, and if so, we secondly pick the new location following the probability density implied by the differential cross-section. The first step consists of computing an interaction probability and drawing a random number to decide whether the pair of numerical particles under consideration interacts. The second step is to randomly choose the scattering angles and update the velocities of the particles.
When the number of numerical particles is sufficiently large, this can provide a good representation of the phase-space and how it evolves under the influence of self-interactions.

For the first step, the probability that two particles interact can be expressed as,\footnotemark\setcounter{FootnoteA}{\value{footnote}}
\footnotetext[\value{FootnoteA}]{We note that Eqs.~\eqref{eq:scatter_probability} and~\eqref{eq:drag_force} can be viewed as a special case of the more general description presented by \cite{Fischer_2025a}. For a derivation, we refer the reader to appendix~A of that paper (the Equations can be obtained by setting $f_\mathrm{bary}=1$ and $\mu=1$).}
\begin{equation} \label{eq:scatter_probability}
    P_{ij} = \frac{\sigma}{m_i} \, m_{\mathrm{n},i} \, |\Delta \mathbf{v}_{ij}| \,\Lambda_{ij} \, \Delta t \,.
\end{equation}
Here, $\Delta \mathbf{v}_{ij}$ is the relative velocity of the particles $i$ and $j$, and $\Delta t$ denotes the time step. 
Moreover, we employ the total cross-section $\sigma$, given by Eq.~\eqref{eq:total_cross_section}. The physical DM particle mass of the species represented by particle $i$ is $m_i$. Whereas the numerical particle mass is $m_{\mathrm{n},i}$.
Importantly, the interaction probability is symmetric, $P_{ij} = P_{ji}$, and the time step should be kept small enough such that $P_{ij} \ll 1$ \citep[but see also appendix~G by][]{Arido_2025}.
A pseudo-random number $x \in [0,1]$ is drawn, and when $x < P_{ij}$, the particles interact. 

If so, in a second step, the post-scattered velocities need to be computed. Therefore, the scattering angles need to be determined. These are the polar angle, $\theta_\mathrm{cms} \in [0, \uppi]$, which must be drawn from the differential cross-section, and the azimuthal angle, $\varphi_\mathrm{cms} \in [0, 2 \uppi]$, which is uniformly distributed over the possible range. These two angles define the scattering direction $\mathbf{e}_\mathrm{scatter}$ (a normalised vector), and the post-scattered velocity of particle $i$ is
\begin{equation}
    \mathbf{v}'_i = \frac{m_j}{m_i + m_j} \, \left( \frac{m_i}{m_j} \, \mathbf{v}_i + \mathbf{v}_j + |\mathbf{v}_i - \mathbf{v}_j| \, \mathbf{e}_\mathrm{scatter} \right) \,.
\end{equation}
After the velocities of the corresponding particle have been updated, and only then, we consider the next pair that contains one of the two particles. For the next pair, we then compute the interaction probability based on the updated velocities. Updating the velocities within a time step is important to be able to conserve energy explicitly when a particle scatters more than once per time step. 

Last, we note that as a consequence of the physical particles represented by a numerical particle not staying close to each other in phase-space due to the self-interactions, an SIDM acceleration of a numerical DM particle is generally undefined.\footnote{This is based on the assumption that we do not resolve the interaction length of the scattering in the simulation, but solve the Boltzmann equation including self-gravity (Eq.~\eqref{eq:vlasov_poisson}). When, instead, the actual scattering process is resolved, this statement does not apply.} There are only discrete velocity kicks, but no gradient of the velocity. This ``gradient-free'' formulation leads to substantial complications compared to other gradient-based schemes employed in $N$-body codes, such as the ones for gravity. This is in particular true when it comes to the parallelisation as we discuss in Sect.~\ref{sec:parallelisation}. But it also implies limitations for the time integration schemes that can be used (see Sect.~\ref{sec:discussion}).

\subsection{Frequent self-interactions} \label{sec:freq_scatter}

\begin{figure}
    \centering
    \includegraphics[width=\columnwidth]{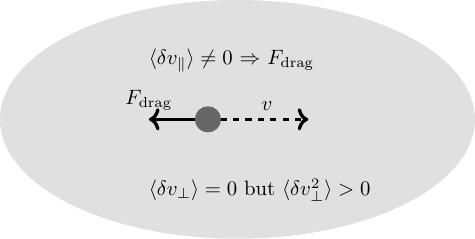}
    \caption{Illustration of the drag force. A phase-space patch of physical DM particles (dark grey) moves through a DM background density. Due to small-angle scatterings, its velocity changes. Each scattering event gives rise to a parallel and perpendicular velocity change compared to the initial direction of motion. Over many scattering events, the parallel velocity changes sum up ($\delta v_\parallel$) and give rise to a drag force ($F_\mathrm{drag}$). The corresponding perpendicular velocity changes are expected to average out ($\delta v_\perp$), i.e.\ the direction of motion does not change. However, the second moment of the perpendicular velocity component increases over time ($\langle \delta v_\perp^2 \rangle > 0$), i.e.\ the perpendicular velocity dispersion of the phase-space patch increases.}
    \label{fig:drag_force}
\end{figure}

Although the scheme described in the previous section can, in principle, simulate any angular dependence, it faces a severe challenge when the scattering rate becomes very large. In order to keep the interaction probability small in that case, it requires tiny time steps. Unfortunately, this does not necessarily imply that the self-interactions are strong enough that one could apply a fluid description. In the case of small-angle scattering, each scattering event has only a small effect, and the overall influence of the self-interactions might be moderate only. Since the scheme of rare scatterings requires tiny time steps, a different formulation is necessary.  

Strongly forward-dominated cross-sections, which are characterised by small scattering angles, can be simulated by employing an effective description derived from first principles \citep{Fischer_2021a}. 
This effective description is based on a drag force decelerating the two particles \citep[see also][]{Kahlhoefer_2014} and perpendicular momentum diffusion. We illustrate this in Fig.~\ref{fig:drag_force}.
The drag force acting on two numerical particles that arise in the limit of Eq.~\eqref{eq:cross-section_fwd_indist} can be expressed as,\footnotemark[\value{FootnoteA}]
\begin{equation} \label{eq:drag_force}
    F_{\mathrm{drag},ij} = \frac{1}{1 + r} \frac{\sigma_\mathrm{\Tilde{T}}}{m_i} \, m_{\mathrm{n},i} \,  m_{\mathrm{n},j} \, |\Delta \mathbf{v}_{ij}|^2 \, \Lambda_{ij} \,.
\end{equation}
Here, $\sigma_\mathrm{\Tilde{T}}$ denotes the modified momentum transfer cross-section as given by Eq.~\eqref{eq:mod_momentum_transfer_cross_section}. In the case of distinguishable particles (Eq.~\eqref{eq:cross-section_fwd_dist}), it must be replaced by the transfer cross-section (Eq.~\eqref{eq:momentum_transfer_cross_section}). The mass of the physical particle represented by particle $i$ is $m_i$. The numerical particle masses of the interacting particles are $m_{\mathrm{n},i}$ and $m_{\mathrm{n},j}$. Their relative velocity is $\Delta\mathbf{v}_{ij}$ and the kernel overlap $\Lambda_{ij}$ is given by Eq.~\eqref{eq:kernel_overlap}.
We employ the mass ratio $r = m_j / m_i$ of the physical particles.
In the case of a single species SIDM model, we have $m_{i} = m_{j}$ and $m_{\mathrm{n},i} = m_{\mathrm{n},j}$.
Moreover, the Equation implies a momentum change, $|\Delta \mathbf{p}_\mathrm{drag}| = F_\mathrm{drag} \, \Delta t$ over the time step $\Delta t$.

The momentum diffusion is modelled by adding the energy lost due to the drag force in a direction perpendicular to the motion of the particles in the centre-of-mass frame. The added momentum is given by
\begin{equation}
    |\Delta \mathbf{p}_\mathrm{rand}| = \sqrt{|\Delta \mathbf{p}_\mathrm{drag}|\left(|\Delta \mathbf{p}_{ij}| - |\Delta \mathbf{p}_\mathrm{drag}| \right)} \, .
\end{equation}
Here we use $\Delta \mathbf{p}_{ij} = m_{\mathrm{n},i} \, \mathbf{v}_i - m_{\mathrm{n},j} \, \mathbf{v}_j$. The momentum $|\Delta \mathbf{p}_\mathrm{rand}|$ is added to each of the two particles but with opposite directions to fulfil momentum conservation. The exact direction of $\mathbf{p}_\mathrm{rand}$ is always perpendicular to the motion in the centre-of-mass frame, but determined by a random angle uniformly distributed in the range $[0, 2 \uppi]$.

Although we have described here the modelling of frequent self-interactions by a drag force and a perpendicular momentum diffusion, it can also be understood as scattering by an effective angle. The underlying physical particles that are represented by the numerical particle would not all have the same deflection angle. Instead, their distribution of deflection angles follows from Molière’s theory \citep{Moliere_1948} for the limit that is relevant here.\footnote{We note that Molière’s theory was formulated for particles scattering off a fixed target, i.e.\ the recoil of the target particles is neglected. Given that we are interested in the regime where the particle's velocities change only slightly, Molière’s theory is applicable when taking care of the reference frame.} In principle, we could sample deflection angles from this distribution. However, this is unnecessarily complicated for the case of elastic scattering. Using an effective angle as a representation of the whole distribution is fine, as the process of stochastic interactions between the numerical particles converges against the correct distribution of deflection angles \citep{Fischer_2021a, Arido_2025}. We want to emphasise that every particle interacts with many particles per time step, namely with all particles for which $\Lambda_{ij} > 0$.

Lastly, we want to emphasise that the description above is consistent with the physical model as expressed by Eq.~\eqref{eq:cross-section_fwd_indist}. This is true independent of the mean-free path. The drag force derived by \cite{Ramos_2025} is incompatible with the numerical representation that we assume (see Sect.~\ref{sec:numerical_representation}). This applies to DSMC codes as well and in particular concerns the second assumption of the derivation by \cite{Ramos_2025}. However, the collective behaviour of our numerical particles would satisfy their drag force. This is also expected to be the case for the more general drag force by \cite{Dvorkin_2014, Munoz_2015} derived in the context of DM-baryon scattering when applied to the SIDM case.

\subsection{Unequal-mass scattering}
For the explanations given so far, we already took the general case of particle scattering with unequal masses into account. Here, we elaborate in greater detail on aspects of unequal-mass scattering.

In Sects.~\ref{sec:rare_scatter} and~\ref{sec:freq_scatter}, we have assumed that the mass ratio of the numerical particle masses is the same as that of the physical particle masses. 
The formulation of the scattering does not directly imply a limit on the physical mass ratio that can be studied. However, for large numerical mass ratios, it can be computationally costly to achieve a sufficiently high resolution, avoiding artificial relaxation from modelling the gravitational interactions. This implicitly limits the parameter space that can be studied. An important characteristic of models with scatterings between particles of unequal masses is that they can give rise to mass segregation, where heavier particles tend to sink towards the centre of DM halos. This can lead to a distinct phenomenology compared to models with equal particle masses \citep[e.g.][]{Patil_2025, Yang_2025a}.

For very large physical mass ratio, it may be advantageous to choose the ratio of masses of the numerical particles different from the ratio of physical masses for the two species. However, this is non-trivial. In particular, \cite{Koda_2011} pointed out that there is no reason to symmetrise the scattering probability for a pair-wise interaction with numerically unequal particle masses to simulate equal-mass scattering. Such a symmetrisation would artificially enhance the scattering rate for lighter particles and reduce it for heavier ones.\footnote{This, for example, implies that the symmetrisation given by eq.~6 in \cite{Rocha_2013} holds only in the case where the mass ratio of the numerical particles is the same as that of the physical ones. However, in this case, their two probabilities used in the symmetrisation are equal by construction as a consequence of their use of a symmetric geometric factor, as we do too (Eq.~\eqref{eq:kernel_overlap}).}

Moreover, obtaining the correct scattering kinematics is not straightforward. When the numerical particles have the same mass ratio as the physical ones, their centre-of-mass velocity is the same. But if the mass ratios do not match, a pair of numerical particles has a different centre of mass velocity than a pair of the physical particles they represent. As a consequence, altering the direction of the momentum vectors of the numerical particles in their centre-of-mass system does not reflect the scattering kinematics of the physical particles they represent. This has a severe consequence: In general, scattering leads to energy equipartition; in the equilibrium state, the temperatures of particles with different masses depend on the mass ratio. If the interactions between the numerical particles do not reflect the scattering kinematics of the physical particles, the simulated system may approach the wrong equilibrium state.
Nevertheless, one may modify the equations that we described above to obtain the correct scattering kinematics in the case of non-matching mass ratios, but this may only be possible when compromising on momentum and/or energy conservation.

As a consequence, previous studies have only considered the case where the numerical mass ratios matched the physical ones. There are only a few studies with $N$-body simulations on unequal-mass scattering. These are \cite{Yang_2025a, Yang_2025b} and \cite{Patil_2025}, where the latter is based on \textsc{OpenGadget3}. They all have in common that they assume a model with two species that differ in the DM particle mass.
Extending the implementation in \textsc{OpenGadget3} beyond the two-species case is left for future work.

Currently, we set the mass ratio of the physical DM particles implicitly with the mass ratio of the numerical particles in the initial conditions (ICs). An additional complication arises for specifying the strength of the interactions. In the two-species case, we assume that the cross-section per mass refers to the lighter species. This is only a conventional choice, and the interaction strength could, in principle, also be specified using the particle mass of the heavier species while the physics remains unchanged.

\subsection{Velocity dependence} \label{sec:velocity_dependence}

Many particle physics models that fall into the class of SIDM have a velocity-dependent cross-section \citep[e.g.][]{Feng_2009, Buckley_2010, Feng_2010, Loeb_2011, Bringmann_2017, Chu_2019, Tsai_2022}. In \textsc{OpenGadget3} we have implemented four different velocity-dependencies that can be used either with isotropic scattering or the forward-dominated limit (Eqs.~\eqref{eq:cross-section_fwd_indist} and~\eqref{eq:cross-section_fwd_dist}).

\begin{figure}
    \centering
    \includegraphics[width=\columnwidth]{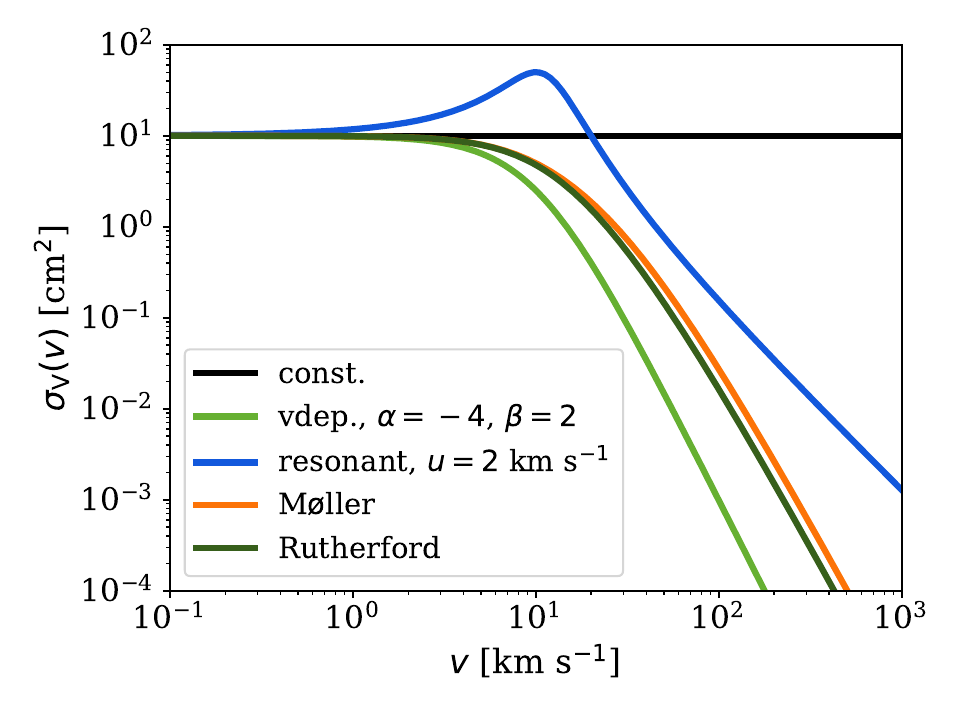}
    \caption{Viscosity cross-section as a function of velocity for different models. We illustrate the velocity dependence of different models implemented in \textsc{OpenGadget3}. These are a velocity-independent cross-section (black), the velocity-dependent model according to Eq.~\eqref{eq:veldep} (light green), a model with a simple parametrisation for a resonant feature as given by Eq.~\eqref{eq:resonant} (blue), M{\o}ller scattering as expressed by Eq.~\eqref{eq:moeller_dcs} (orange), and Rutherford scattering following Eq.~\eqref{eq:rutherford_dcs} (dark green).
    The cross-section normalisation parameter is set to $\sigma_{0,\mathrm{V}} = 10 \, \mathrm{cm}^2 \, \mathrm{g}^{-1}$ and for the velocity-dependent cross-section, $w = 10 \, \mathrm{km} \, \mathrm{s}^{-1}$.
    }
    \label{fig:vdep_cross-sections}
\end{figure}

The first model parametrises the velocity dependence of the angle-averaged cross-section, $X$, as
\begin{equation} \label{eq:veldep}
    \sigma_X= \sigma_{0,X} \, \left(1+ \left( \frac{v}{w} \right)^\beta \right)^{\alpha/\beta} \,.
\end{equation}
Here, the cross-section is constant for $v \ll w$ and decreases as $\sigma_X \propto v^{\alpha}$ for $v \gg w$.  The transition between these two regimes is governed by the parameter $\beta$. The parameters should be chosen such that they fulfil $w > 0$, $\alpha < -1$, and $\beta > 0$ to fall in the range currently supported by the SIDM time step criterion (see Sect.~\ref{sec:time_step_criterion}).

In addition, we include a model with a resonance, i.e.\ the cross-section is enhanced for velocities close to $w$. This is a feature expected from various particle physics models \citep{Tulin_2013a, Chu_2019, Chua_2020}. In detail, we use the same formulation as \cite{Sanchez_2025}, namely
\begin{equation} \label{eq:resonant}
    \sigma_X = \sigma_{0,X} \, \frac{1 + w^2/u^2}{1 + [v - w]^2 / u^2} \,,
\end{equation}
where $u$ is an additional parameter that determines the width and height of the resonant peak. We leave it for future work to implement more complicated models as done by \cite{Tran_2024}.

Moreover, we have included the velocity dependence of M{\o}ller scattering. In detail, the one of the viscosity cross-section (Eq.~\eqref{eq:viscosity_cross_section}).
\begin{multline} \label{eq:moeller_viscosity}
    \sigma_\mathrm{V} = \frac{6 \, \sigma_{0,\mathrm{V}} \, w^8}{v^8 + 2 \, v^6 \, w^2} \left[ 2 \, \left(5+\frac{5 \, v^2}{w^2}+\frac{v^4}{w^4}\right) \, \ln\left(1+\frac{v^2}{w^2}\right) \right. \\
    \left. - 5 \, \left( \frac{2 \, v^2}{w^2} + \frac{v^4}{w^4}\right) \right] \,.
\end{multline}
It is normalised such that $\sigma_\mathrm{V}(v=0) = \sigma_{0,\mathrm{V}}$.
Furthermore, the velocity dependence of Rutherford scattering is available in \textsc{OpenGadget3}. 
\begin{equation} \label{eq:rutherford_viscosity}
    \sigma_\mathrm{V} = \frac{6 \, \sigma_{0,\mathrm{V}} \, w^6}{v^6} \left[ \left(2+\frac{v^2}{w^2}\right) \, \ln\left(1+\frac{v^2}{w^2}\right) - \frac{2 \, v^2}{w^2}\right]
\end{equation}
As for M{\o}ller scattering, we obtained it by computing the viscosity cross-section and normalised it such that $\sigma_\mathrm{V}(v=0) = \sigma_{0,\mathrm{V}}$.
A derivation and discussion of these velocity-dependencies is, for example, given by \cite{Girmohanta_2022}.
While these velocity dependencies do not capture the angular dependence, we note that we have also implemented the full differential cross-sections for M{\o}ller and Rutherford scattering as explained in Sect.~\ref{sec:differential_cross_section}.

The velocity dependencies from above can be used together with isotropic scattering or the forward-dominated limit. The underlying idea is to assume that the angular and velocity dependences are separable. Although this typically does not apply to realistic models, it can be helpful for understanding the role of angle and velocity dependence to define the cross-section as
\begin{equation} \label{eq:decoupled_dependences}
    \frac{\mathrm{d}\sigma}{\mathrm{d}\cos\theta_{\mathrm{cms}}} = f_\mathrm{angle}(\theta) \, f_\mathrm{vel}(v) \,.
\end{equation}
In the isotropic case, $f_\mathrm{angle}$ would be a constant, $\sigma_0$, and in the forward-dominated case it would be given by Eq.~\eqref{eq:cross-section_fwd_indist} or Eq.~\eqref{eq:cross-section_fwd_dist}. Similarly, in the case of a velocity-independent cross-section, $f_\mathrm{vel}(v)$ would be unity. Alternatively, it could be given by $\sigma_X / \sigma_{0,X}$ using one of the Eqs.~\eqref{eq:veldep}--\eqref{eq:rutherford_viscosity}.

\subsection{Full differential cross-sections} \label{sec:differential_cross_section}

Many particle physics models feature interactions for which the velocity and angular dependences cannot be viewed separately, i.e.\ the differential cross-section cannot be expressed in a form like Eq.~\eqref{eq:decoupled_dependences}.
Simulating such SIDM cross-sections was first explored by \cite{Robertson_2017b} and later implemented in several codes.
\textsc{OpenGadget3} allows the simulation of full differential cross-sections too. It was initially implemented by \cite{Wiertel_2023}, and we will describe it in the following.

For generality, we assume to model the differential cross-section in the range $\theta_\mathrm{cms} \in [\theta_\mathrm{min}, \theta_\mathrm{max}]$. This allows us to exclude problematic angular ranges, e.g.\ for divergent models, as also done by \cite{Robertson_2017b}. The full angular range would be $\theta_\mathrm{min} = 0$ and $\theta_\mathrm{max} = \uppi$.
To compute the interaction probability (Eq.~\eqref{eq:scatter_probability}), we need to compute the total cross-section for the corresponding angular range,
\begin{equation} \label{eq:total_cross_section_mod}
    \sigma_{[\theta_\mathrm{min},\, \theta_\mathrm{max}]}(v) = \int_{\theta_\mathrm{min}}^{\theta_\mathrm{max}}\frac{\mathrm{d\sigma}(\theta', v)}{\mathrm{d}\cos \theta_\mathrm{cms}} \, \sin(\theta') \, \mathrm{d}\theta' \,.
\end{equation}
At the beginning of the simulation, we compute $\sigma_{[\theta_\mathrm{min},\, \theta_\mathrm{max}]}$ for various values of $v$ by numerically integrating the differential cross-section. These values are stored in a lookup table, which is used to interpolate the total cross-section at the requested velocity.

To compute the scattering angle, we compute the cumulative distribution function for the scattering angles,
\begin{align} \label{eq:cdf}
    \mathrm{CDF}(\theta_\mathrm{cms}, v) = &\frac{1}{\sigma_{[\theta_\mathrm{min},\, \theta_\mathrm{max}]}(v)} \nonumber \\ &\int_{\theta_\mathrm{min}}^{\theta_\mathrm{cms}} \frac{\mathrm{d\sigma}(\theta', v)}{\mathrm{d}\cos\theta_\mathrm{cms}} \, \sin(\theta') \, \mathrm{d}\theta' \,.
\end{align}
The CDF increases monotonically from $0$ to $1$ and gives the probability that if the particles interact, they scatter with an angle smaller than $\theta_\mathrm{cms}$. To determine the scattering angle $\theta_\mathrm{cms}$, we draw a uniformly distributed random number $x \in [0,1]$ and obtain $\theta_\mathrm{cms}$ by inverting $x = \mathrm{CDF}(\theta_\mathrm{cms}, v)$.

The scattering angle is determined from a lookup table by interpolation. 
In \textsc{OpenGadget3}, two different methods are implemented to do so. Here, we briefly describe them; a more detailed description can be found in Appendix~\ref{sec:sampling_differential_cross_section}.

The first one is the Grid method, which uses a grid in the $x$--$v$ plane and stores the values for the scattering angle. The grid is linearly spaced in $x$ and allows us to easily infer the scattering angles by interpolating between the grid points. 
However, computing the lookup table is a little more complicated, as the corresponding values of $\theta_\mathrm{cms}$ must be found for the given values of $x$.

The second method implemented in \textsc{OpenGadget3} is the ZSlice method. Here, a grid in the $\theta_\mathrm{cms}$--$v$ plane is used to store the values of $x$. The table can be directly computed using Eq.~\eqref{eq:cdf}. However, inferring the scattering angle from it is computationally more expensive than for the Grid method. But compared to the Grid method, the ZSlice method is able to significantly reduce the numerical error in the most error-prone regions and thus yields overall a higher accuracy when using the same number of grid points. At the same time, no substantial slowdown of the simulation was found due to the increased computational costs in evaluating the scattering angle \citep{Wiertel_2023}. Therefore, we take the ZSlice method as the default method. 

The tables to store the cumulative distribution function and the angle-averaged cross-sections are logarithmically spaced in velocity, but additionally contain an entry for a velocity of zero. If interacting particles encounter a relative velocity larger than the maximum value stored in the tables, we automatically extend the tables.

We have implemented M{\o}ller scattering in the non-relativistic limit using the Born approximation.
\begin{equation} \label{eq:moeller_dcs}
    \left.\frac{\mathrm{d}\sigma}{\mathrm{d}\cos\theta_\mathrm{cms}}\right|_\textnormal{M{\o}ller} = \sigma_0 \frac{\left(3 \cos^2\theta_\mathrm{cms} + 1 \right) \frac{v^4}{w^4} + 4 \frac{v^2}{w^2} + 4}{\left( \sin^2\theta_\mathrm{cms} \, \frac{v^4}{w^4} + 4 \frac{v^2}{w^2} + 4 \right)^2} \,.
\end{equation}
Rutherford scattering is also available in \textsc{OpenGadget3}.
\begin{equation} \label{eq:rutherford_dcs}
    \left.\frac{\mathrm{d}\sigma}{\mathrm{d}\cos\theta_\mathrm{cms}}\right|_\textnormal{Rutherford} = \sigma_0 \left[2 + \frac{v^2}{w^2} (1-\cos \theta_\mathrm{cms}) \right]^{-2} \,.
\end{equation}
We note that for these two models, the scattering is isotropic at $v=0$ with a total cross-section (Eq.~\eqref{eq:total_cross_section}) equal to $\sigma_0 /2$.
Further differential cross-sections can be easily added, as all derived quantities, i.e.\ the angle-averaged cross-sections as well as the cumulative distribution function to sample the scattering angles, are computed numerically by the code.

The description above has so far assumed that the scattering is only modelled with the scheme for rare interactions (Sect.~\ref{sec:rare_scatter}). However, the differential cross-section could be strongly anisotropic, such that the rare scattering scheme is hardly applicable because of high computational costs. Instead, one may model the interactions with the scheme for frequent scattering, and approximate the differential cross-section with Eqs.~\eqref{eq:cross-section_fwd_indist} or~\eqref{eq:cross-section_fwd_dist} for the angular dependence and use a separable velocity dependence (Sect.~\ref{sec:velocity_dependence}). 
However, this is only applicable if the differential cross-section does not give rise to significant large-angle scattering.
But if this is the case, one can combine the two schemes into a hybrid SIDM scheme as explored by \cite{Arido_2025}.
This hybrid scheme, as implemented in \textsc{OpenGadget3}, splits the differential cross-section according to a critical angle, $\theta_\mathrm{c}$, into two parts. Scatterings by an angle smaller than $\theta_\mathrm{c}$ are described by the frequent scattering scheme, and scatterings above $\theta_\mathrm{c}$ are modelled with the help of the rare scattering scheme.

Accordingly, the momentum transfer or modified momentum transfer cross-sections relevant for the frequent scheme are integrated over the interval $\theta_\mathrm{cms} \in [0, \theta_\mathrm{c}]$. Analogous to the total cross-section, we use a lookup table from which we interpolate the value at the requested velocity. Those are then used to compute the drag force according to Eq.~\eqref{eq:drag_force}. The integration range for the total cross-section is set to $\theta_\mathrm{cms} \in (\theta_\mathrm{min} = \theta_c, \theta_\mathrm{max} = \uppi]$. 

In the case of models with indistinguishable particles, where the differential cross-section is symmetric (e.g.\ M{\o}ller scattering), it is also favourable to describe scatterings close to $\theta_\mathrm{cms} = \uppi$ with the frequent scheme too. Hence, we integrate the modified momentum transfer cross-section over the joint interval $\theta_\mathrm{cms} \in [0, \theta_\mathrm{c}] \cup [\uppi - \theta_\mathrm{c}, \uppi]$ and the total cross-section over the interval $\theta_\mathrm{cms} \in (\theta_\mathrm{min} = \theta_\mathrm{c}, \theta_\mathrm{max} = \uppi - \theta_\mathrm{c})$. Furthermore, the angular range of the cumulative distribution function to sample the scattering angle, $\theta_\mathrm{cms}$, must be adjusted accordingly. However, in practice, we have two independent critical angles, one for forward scattering and one for backward scattering. 

The choice of the critical angle, $\theta_\mathrm{c}$, is a trade-off between accuracy and a reduction in computational costs. The larger $\theta_\mathrm{c}$, the less accurate the deflection of particles due to small scattering angles is modelled. But at the same time, the simulation can become much faster. How large the speed-up is depends in detail on the anisotropy of the cross-section. In the case of typical astrophysical set-ups such as merging galaxy clusters, satellite galaxies orbiting their host, or full cosmological simulations, our default value of $\theta_\mathrm{c} = 0.1$ may yield sufficiently accurate results.
For more details on this hybrid scheme, we refer the reader to \cite{Arido_2025}.

\subsection{Time integration} \label{sec:time_integration}
In \textsc{OpenGadget3}, the time integration is performed using a leapfrog scheme in the kick-drift-kick (KDK) formulation. This integration scheme is commonly used in cosmological $N$-body codes due to its symplectic nature, and the KDK formulation is particularly helpful for the synchronisation when particles reside on different time steps following a block time step scheme.

The computation during one time step is structured as follows:
\begin{enumerate}
    \item[I.] First, a half-step kick is executed, which means the velocities of the active particles are evolved by half a time step, given their acceleration,\footnote{We are using a notation where the subscript denotes the particle index and the superscript is not an exponent, but refers to the time instance.} 
    \begin{equation}
        \mathbf{v}^{n+1/2}_i = \mathbf{v}^{n}_i + \mathbf{a}^{n}_i \, \Delta t_i / 2 \,.
    \end{equation}
    \item[II.] The drift operation is executed, namely, the particle's positions are updated by one time step,
    \begin{equation}
        \mathbf{x}^{n+1}_i = \mathbf{x}^{n}_i + \mathbf{v}^{n+1/2}_i \, \Delta t_i \,.
    \end{equation}
    \item[III.] Now, the forces acting on the particles resulting in their acceleration are computed. The acceleration of the particles results, for example, from gravity or hydrodynamics,
    \begin{align}
        \mathbf{a}^{n+1}_i = \,& \mathbf{a}^{n+1}_{\mathrm{grav},ij}(\mathbf{x}^{n+1}_j) \nonumber\\
        &+ \mathbf{a}^{n+1}_{\mathrm{hydro},ij}(\mathbf{x}^{n+1}_j, ...) + ...\,.
    \end{align}
    \item[IV.] The pre-scattering velocities\footnote{Here, we use $\mathbf{u}^{n+1}$ to denote the velocity before the velocity kicks due to the self-interactions have been applied. This allows us to distinguish it from the velocity $\mathbf{v}^{n+1}$ after the scattering.} are calculated by completing the KDK scheme with the second half-step kick,
    \begin{equation}
        \mathbf{u}^{n+1}_i = \mathbf{v}^{n+1/2}_i + \mathbf{a}^{n+1}_i \, \Delta t_i / 2\,.
    \end{equation}
    \item[V.] Depending on the physics that are modelled, some corresponding computations, such as for black holes or cosmic rays, might be executed after the second half-step kick.
    \item[VI.] Last, before we return to I.\ for the first half-step kick of the next time step, we compute the DM self-interactions,
    \begin{align}
        \mathbf{v}^{n+1}_i = \mathbf{u}^{n+1}_i + &\mathbf{\Delta v}_\mathrm{scatter}^{n+1}\bigg(\mathbf{x}^{n+1}_j, \nonumber\\
        &\mathbf{u}^{n+1}_j, \Delta t_i, \frac{\mathrm{d}\sigma}{\mathrm{d}\cos\theta_\mathrm{cms}}\bigg) \,.
    \end{align}
\end{enumerate}
In the last step (VI.), we compute the self-interactions between the two half-step kicks. It is not common to all SIDM implementations to compute the velocity kicks resulting from the scattering at this place. For example, the implementation by \cite{Robertson_2017a, Robertson_2017b} (in \textsc{Gadget-3}) performs the scattering between the first half-step kick (I.) and the drift (II.) operations. The implementation in \textsc{Gadget-2} by \cite{Yang_2021} follows the same scheme. For the SIDM implementation in \textsc{GIZMO} \citep[based on][]{Rocha_2013},\footnote{As employed by \citet{Robles_2017}.}  the scattering is computed between the drift (II.) and second kick (III.) operations. Another alternative is the approach implemented in \textsc{SWIFT}, which splits the drift operation (II.) into two half-step drifts and executes the self-interaction in between those drifts \citep[actually it is a little more complicated, see sect.~2.3 by][]{Correa_2022}. Possible differences arising from the choice of the position at which the self-interactions are computed have not yet been investigated thoroughly. However, given that the gravitational force computation does only depend on the positions of the particles, whereas the DM self-interactions only alter the velocities, a problematic interference arising from the SIDM computations appears to be less plausible.

The adaptive time-stepping in \textsc{OpenGadget3} is based on a block time-step scheme. It considers a particle to be active or passive depending on the block time step assigned to the particle and the time step that is being computed. The self-interactions are computed between pairs of two active particles and pairs of one active and one passive particle. As the neighbour search is performed for each active particle, we find the pairs consisting of two active particles twice per time step. In practice, we only perform the computations of the SIDM interactions when the identification number of the first particle of the pair is larger than that of the second particle. This way, we consider each pair only once and use the full time step for the scattering computation.

\subsection{Time step criterion} \label{sec:time_step_criterion}

To ensure accurate simulation results, it is crucial to limit the numerical errors by choosing a sufficiently small time step. This can be achieved by formulating the time-step criterion to limit the interaction probability in the rare-scattering scheme. For the frequent-scattering scheme, the relative velocity change implied by the drag force should be kept small. In the following, we explain the formulation of the SIDM time-step criteria in \textsc{OpenGadget3}.

Starting from Eqs.~\eqref{eq:scatter_probability} and~\eqref{eq:drag_force}, one can derive the time-step criteria. To control the size of the time step, we introduce the parameters $\tau_\mathrm{rare}$ and $\tau_\mathrm{freq}$, which limit the interaction probability and the relative velocity change implied by the drag force, for the two schemes, respectively.
For the rare scattering scheme, the time step criterion is
\begin{equation} \label{eq:time_step_crit_rare}
    \Delta t_\mathrm{rare} = \frac{\tau_\mathrm{rare}}{m_{\mathrm{n},x} \, \Lambda_{ii}} \left[\max_{v \, \in \, [0, \, v_\mathrm{max}]}\left(\frac{\sigma(v)}{m_x} \, v\right)\right]^{-1} \,.
\end{equation}
Analogously, for the frequent scattering scheme, it is
\begin{equation} \label{eq:time_step_crit_freq}
    \Delta t_\mathrm{freq} = \frac{\tau_\mathrm{freq}}{m_{\mathrm{n},x} \, \Lambda_{ii}} \left[\max_{v \, \in \, [0, \, v_\mathrm{max}]}\left(\frac{\sigma_\mathrm{\Tilde{T}}(v)}{m_x} \, v\right)\right]^{-1} \,.
\end{equation}
Some quantities in the equations for the interaction probability and the drag force are subject to substantial noise and depend on which neighbours a particle sees. For example, the maximum relative velocity that occurs for all the pairs of which a particle is part in a given time step may fluctuate. The size of this fluctuation also depends on the number of neighbours used to determine the kernel size.
We aim for a stringent and robust time step criterion. To cope with these fluctuations, we try to estimate the maximum of the kernel overlap integral and the maximum of the product of cross-section and relative velocity. For a discussion on how to build a SIDM time-step criterion, we refer the reader to appendix~B by \cite{Fischer_2024a}.

For the kernel overlap, we use the self-overlap, i.e.\ $\Lambda_{ii}$, in Eqs.~\eqref{eq:time_step_crit_rare} and~\eqref{eq:time_step_crit_freq}. This gives an upper bound on the value that $\Lambda_{ij}$ possibly can take. We note that the use of $\Lambda_{ii}$ is based on the assumption that the set of particles employed in determining the kernel size is the same as for the formation of the interaction pairs. In case this is not applicable, a different formulation could be favourable \cite[see Sect.~2.4.2 by][]{Fischer_2025a}.

Estimating the maximum value for $\sigma_X(v)/m_x \, v$ can be more complicated. Some velocity-dependent cross-sections have a well-defined global maximum that could be used. Nevertheless, we keep track of the maximum relative velocity, $v_\mathrm{max}$, that a particle has seen in the last time step. This allows us to limit the relevant domain within which we want to find the maximum of $\sigma_X(v)/m_x \, v$, as expressed in Eqs.~\eqref{eq:time_step_crit_rare} and~\eqref{eq:time_step_crit_freq}.
In the case of the full differential cross-section implementation, we have a lookup table for this maximum as a function of $v_\mathrm{max}$. But for the velocity-dependences expressed in Sect.~\ref{sec:velocity_dependence}, a different approach is possible.

In this case, there is a global maximum for $\sigma_X(v)/m_x \, v$ at a velocity $v_e$. If $v_\mathrm{max} > v_e$, we use the global maximum, $\sigma_X(v_e)/m_x \, v_e$, for the time-step criterion. Otherwise we employ $\sigma_X(v_\mathrm{max})/m_x \, v_\mathrm{max}$, as $\sigma_X(v)/m_x \, v$ is monotonically increasing for $v < v_e$ in all the cases we consider.
Specifically, $v_e$ takes the following values for the models considered in Sect.~\ref{sec:velocity_dependence}.
In the case of the velocity-dependent model according to Eq.~\eqref{eq:veldep}, it is
\begin{equation}
    v_e = \frac{w}{(-1-\alpha)^{1/\beta}} \,.
\end{equation}
For the model with a resonant feature (Eq.~\eqref{eq:resonant}) we obtain
\begin{equation}
    v_e = \sqrt{w^2 + u^2} \,.
\end{equation}
Moreover, for the viscosity cross-section of M{\o}ller scattering (Eq.~\eqref{eq:moeller_viscosity}) we find the maximum of $\sigma_X(v)/m_x \, v$ at
\begin{equation}
    v_e \approx 0.9845 \, w \,,
\end{equation}
and in the case of Rutherford scattering (Eq.~\eqref{eq:rutherford_viscosity}) it is
\begin{equation}
    v_e \approx 0.8996 \, w \,.
\end{equation}

With the scattering computations, we keep track of the maximum velocity that a particle sees in a time step.
However, at the beginning of a simulation, no scattering has been computed yet, and this information is missing. Therefore, we perform additional computations to go over all pairs and find the maximum relative velocity for every particle.  

In \textsc{OpenGadget3}, the time step is set based on the minimum of all time-step criteria. For a simulation with self-interactions only, this would be
\begin{equation}
    \Delta t_\mathrm{SIDM} = \min(\Delta t_\mathrm{freq}, \Delta t_\mathrm{rare}) \,.
\end{equation}
We note that the time-step criteria do not directly give the time step at which a particle is evolved; instead, a time-stepping function assigns it to a block time step. These block time steps are given by a base time step multiplied by a power of two.
Thus, the time-step criteria specify an upper bound, with the actual assigned block time step being the largest that does not exceed that bound, which can therefore be smaller by up to a factor of two.

\subsection{Comoving Integration}
To run cosmological simulations, we use comoving integration in \textsc{OpenGadget3}. This means that the expanding space is taken into account as a uniform background effect by formulating the computations based on comoving coordinates. Moreover, instead of peculiar velocities, the canonical momentum, $\hat{v} = a \, v$, related through the scale factor, $a$, is used.
As a consequence, additional factors are introduced that are taken into account when computing the time step.
This allows us to properly account for their dependence on the scale factor.

The time step, as computed in \textsc{OpenGadget3}, for integrating from the time instance $n$ to $n+1$ is
\begin{equation}
    \Delta \hat{t}^*_\mathrm{SIDM} = \int^{a^{n+1}}_{a^n} \frac{1}{\hat{H}(a) \, a^5} \, \mathrm{d}a \,.
\end{equation}
Here, $a$ denotes the scale factor\footnote{The superscript in the expressions $a^n$ and $a^{n+1}$ is not an exponent but an indication of the time instance that we refer to.} and $\hat{H} = H/h$ with $H$ being the Hubble parameter, and the implicit definition of $h$ via the Hubble constant, $H_0 = h \times 100 \, \mathrm{km} \, \mathrm{s}^{-1} \, \mathrm{Mpc}^{-1}$.
In addition, we modify the time step criterion for comoving integration to take into account the dependence on the scale factor. The time step criterion is expressed in terms of $\eta = \log(a)$.
In the case of the scheme for rare interactions, the time step criterion (Eq.~\eqref{eq:time_step_crit_rare}) becomes
\begin{equation} \label{eq:time_step_crit_rare_como}
    \Delta \eta_\mathrm{rare} = \frac{\tau_\mathrm{rare}}{\hat{m}_{\mathrm{n},x} \, \hat{\Lambda}_{ii}} \left[\max_{\hat{v} \, \in \, [0, \, \hat{v}_\mathrm{max}]}\left(\frac{\sigma(v)}{m_x} \, \hat{v}\right)\right]^{-1} \, \frac{a^4 \, \hat{H}(a)}{h}\,.
\end{equation}
Here we use $\hat{m}_{\mathrm{n},x} = m_{\mathrm{n},x} \, h$ and $\hat{\Lambda}_{ii} = \Lambda_{ii} \, h^3 / a^3$. The case of frequent interactions (Eq.~\eqref{eq:time_step_crit_freq}) is analogous to the rare interaction scheme. 

\subsection{Implementation overview} \label{sec:implementation_overview}

\begin{figure*}
    \centering
    \includegraphics[width=0.75\linewidth]{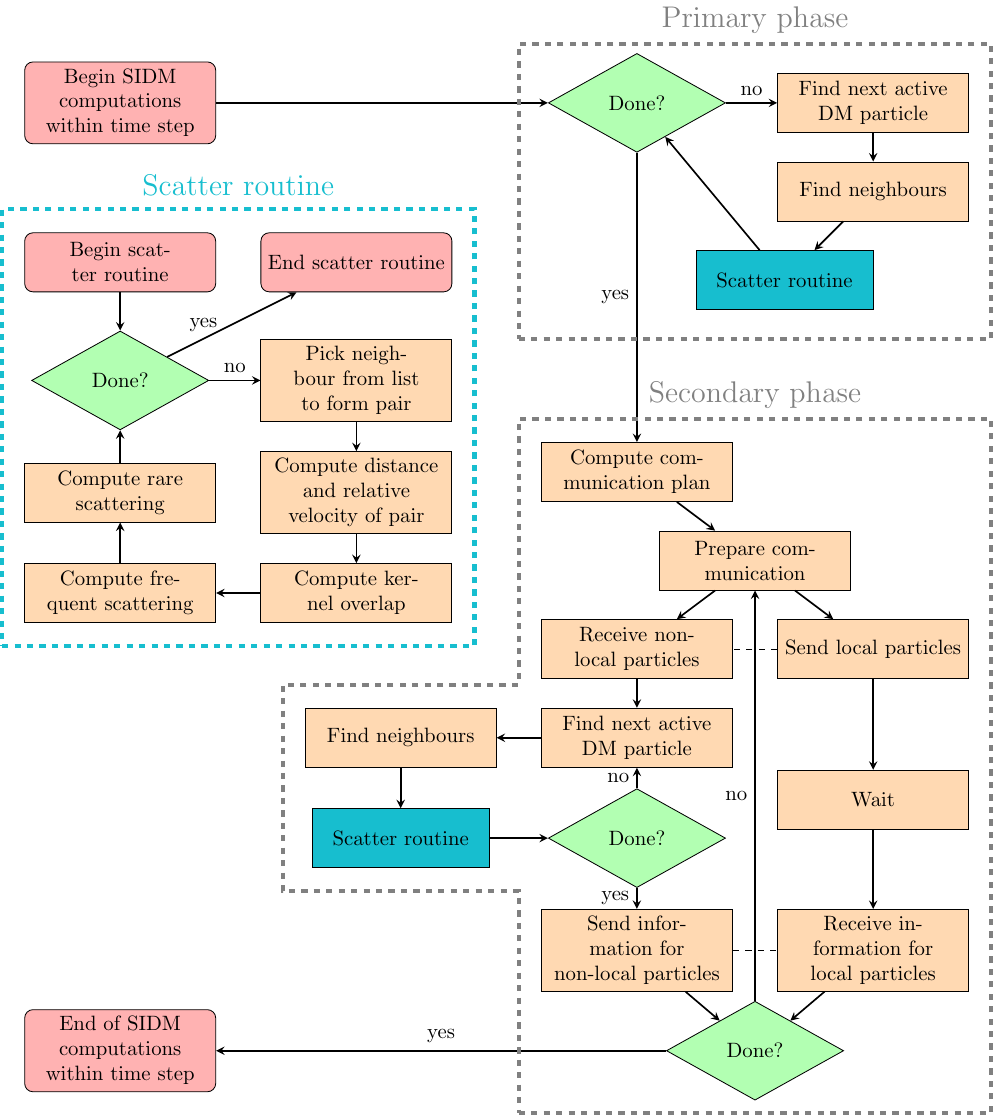}
    \caption{Flow chart for our implementation of SIDM in \textsc{OpenGadget3}. Only the part for the scattering that is executed every time step is shown. The dashed blue box gives the scatter routine, which is executed for every pair as part of the primary and secondary phases.}
    \label{fig:flowchart}
\end{figure*}

The main computations for SIDM are executed within the loop that iterates over the time steps. As already described in Sect.~\ref{sec:time_integration}, we compute the scattering after the second half-step kick. Here, \textsc{OpenGadget3} also computes other non-standard physics. Except for computations that are needed when starting a run, e.g.\ computing lookup tables; only the time step criteria and kernel size computations are performed elsewhere.

The time step criteria are evaluated at the beginning of the time step, i.e.\ before the first half step kick.
The kernel size is determined at a later stage,
namely after the computation of the gravitational accelerations and before the computation of the hydrodynamical accelerations, following the pre-existing structure of \textsc{OpenGadget3}. However, this has the disadvantage that the kernel sizes can change after the time step has been set. This makes the time step criterion a little less robust. An alternative would be to determine the kernel sizes even before the time step computation. Given that the particles are drifted afterward, the actual number of neighbours for the scattering computations could vary. We have not explored and evaluated this option.

In Fig.~\ref{fig:flowchart}, we provide a flow chart for the computations to model the DM scattering itself, i.e.\ the ones that happen after the second half-step kick.
Here, the interactions for the local particles are computed first in the primary phase. Afterwards, the pairs for which the MPI processes need to exchange information are handled in the secondary phase.
Overall, this is only a fraction of the flow chart that \textsc{OpenGadget3} follows. For an overview of all computations, we refer the reader to \cite{Dolag_2026}.

The SIDM module in \textsc{OpenGadget3} is written in C++17. Although it is written as C++ code, it contains a few compromises to connect to other C-style parts of the code.
The module contains many options, e.g.\ in terms of particle physics models that can be simulated. To allow for this without repeating a lot of code and achieving it at compile time, parts of the SIDM module are expressed in an abstract manner, making use of the template metaprogramming capabilities of C++. This is also meant to enhance the maintainability of the code.

In the following, we focus on the parallelisation of the SIDM computations as this is an intrinsically difficult task. Many other parts of the code are described in the main \textsc{OpenGadget3} release paper and the references therein \citep{Dolag_2026}.

\subsection{Parallelisation} \label{sec:parallelisation}
In the following, we describe the parallelisation strategy for the SIDM module in \textsc{OpenGadget3}.
It consists of two parts: the parallelisation for distributed memory to facilitate computations across computing nodes, and the shared memory parallelisation. The first one is realised by employing the MPI, and the second one with the help of OpenMP.

Parallelising the computations for SIDM is more challenging than it is typically for gravity or hydrodynamics. This has to do with the fact that SIDM is formulated in a gradient-free fashion; in contrast to other physics modules, we are not computing an acceleration for every particle, but a discrete velocity kick. To ensure the explicit conservation of energy, it is necessary to apply this kick immediately. The kicks are computed on a pairwise basis and alter the velocities of the two particles belonging to the pair. During a time step, a particle is typically part of multiple pairs formed with its neighbours. If a particle interacts more than once per time step, it is important that the velocity for the second and subsequent interactions is updated; using the velocity from the beginning of the time step would violate energy conservation. We note that the interaction probability or drag force should be computed from the updated velocities, as they depend on the velocity. This is necessary to be consistent with the scattering physics.

If computations are executed in serial, this is not much of a problem, as all pairs are computed consecutively. However, for parallel computations, this becomes much more complicated; one needs to figure out which pairs can be safely computed in parallel.
We start by describing the MPI parallelisation and continue with the OpenMP parallelisation.

\subsubsection{MPI parallelisation}
Multiple processes are launched with the start of a simulation, and do not have access to the same memory. The simulation particles are distributed over the processes and, with that, over the available memory. While in practice every process (also known as MPI rank) first executes the SIDM computations among the local particles, i.e.\ the ones it has access to, the second phase for interactions of particles that do not belong to the same process requires communication among the processes to exchange the relevant information.

\begin{figure}
    \centering
    \includegraphics[width=0.75\columnwidth]{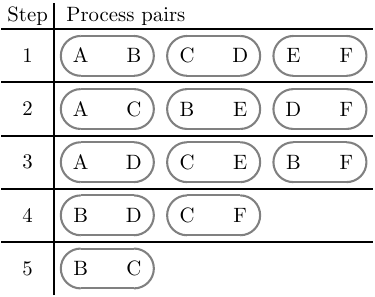}
    \caption{An example of a communication plan that is used to facilitate communication between the processes (MPI ranks). The steps are executed in a consecutive manner; only disjoint pairs can be assigned to a step. We note that in our example, no communication is required for the pairs: A-E, A-F, and D-E.}
    \label{fig:communication_plan}
\end{figure}

The first phase is unproblematic for the MPI parallelisation, and we will now explain how the second phase is realised. The MPI parallelisation of our SIDM module has also been previously described by \cite{Fischer_2024a}.
For the non-local SIDM computations, particles are sent to other MPI ranks to do the interactions and their updated properties, in particular, their velocity is sent back. Here, it is important that a particle is not sent to multiple processes at a time or sent at all if needed locally for particles that are received from another process.
The general idea for parallelisation is to find pairs of processes for which we can execute the non-local computations in parallel.
The non-local computations are executed in multiple steps, where for each step only pairs that do not conflict with each other are executed. When all computations and communications for a specific step are completed, the next step with a new set of pairs of processes can start.
The complications lie in building the communication plan, i.e.\ assigning the process pairs to the communication steps. This plan tells every process when it has to communicate with which process, an example for such a plan is given in Fig.~\ref{fig:communication_plan}.
In practice, we build the communication schedule for every process following seven steps: 
\begin{enumerate}
    \item Given that every process knows how many particles it will send to and receive from other processes, it creates a list with all its communications to other processes. The list also contains the number of particles that are exchanged with each process (the sum of send and receive). It is used later on to prioritise the communication pairs of processes.
    \item Each process sorts the previously created list according to the number of particles to be exchanged, i.e.\ according to priority.
    \item Next, all processes exchange their communication lists. As a result, every process knows about all communications that should be executed. We note that the length of these lists need not scale as $N^2$ ($N$ being the number of processes). In cases where communication is not required between all possible pairs of processes (which is typically the case), it can be shorter.
    \item From the received lists, every process builds a single list with all communications. Here, the processes start with the highest priority elements of all individual lists and keep for each communication pair of processes only one entry.
    Moreover, the final list contains only communication pairs for which a non-zero number of particles is exchanged.
    \item The list containing all communications from the previous step is sorted according to priority. We note that we don't require that the list be perfectly sorted. Instead, we interrupt the sorting when the list is sufficiently ordered. In practice, the list is split in two parts, a sorted one with the highest priorities and an unsorted part of length $N_\mathrm{unsrt}$ with priorities $p_i$. We interrupt the sorting once
    \begin{equation}
    \sqrt{N_\mathrm{unsrt} \frac{\sum_i^{N_\mathrm{unsrt}} p_i^2}{\left(\sum_i^{N_\mathrm{unsrt}} p_i\right)^2} - 1} \, < \, 0.7
    \end{equation}
    is fulfilled for the unsorted part of the list.
    
    \item Based on the partially sorted list, every process builds the overall communication plan on its own; no further communication is involved. The plan consists of several steps that are meant to be executed in a consecutive order. At each step, several pairs do their communication and also the scattering computations of the exchanged particles. These pairs do not conflict with each other, i.e., no process can be assigned more than once per step. The plan is built by trying to fulfil the highest priorities first. The steps are filled after each other with pair-wise communications. If a communication does not fit in a step because a corresponding process is already busy, it is queued and retried for the next step. We turn to the next step when the queue for the communication that should be assigned for the current step is empty or when all processes are already assigned a communication for the current step. An example of the communication plan is given in Fig.~\ref{fig:communication_plan}.
    \item Last, every process extracts its communication schedule from the overall communication plan.
\end{enumerate}
When every process knows its communication schedule, the communications and SIDM computations of the second phase can start.

We note that this parallelisation scheme implies significant waiting time. For example, for each pair of processes, only one of the two processes is executing the SIDM computations at a time, while the other one is waiting.
Moreover, all pairs of a step need to complete their computations first before proceeding with the next step of the communication plan. The main reason to prioritise processor pairs according to the number of particles they exchange is to reduce this waiting time.

\subsubsection{OpenMP parallelisation}

The computations executed by an MPI rank can be further parallelized with OpenMP.
Therefore, multiple threads are created that all have access to the same memory. For every particle, we have a list of neighbours, i.e.\ a list of particles to form pairs with for the computation of the self-interactions. When thread A is processing particle $i$ with its neighbour list containing particle $j$, it eventually alters the properties of particles $i$ and $j$. This implies that at the same time, no other thread can be allowed to alter the properties of particles $i$ and $j$. Instead, thread B would need to wait for A to finish with the pair $ij$ before it can compute an interaction based on the velocity of $i$ or $j$. Within OpenMP, we realise this by using locks, i.e.\ for every particle, there exists a lock that a thread closes when it starts to process that particle and that it opens again when it has completed processing it. In order to save waiting time, thread B would not wait until it can process particle $i$, but instead continue with the next particle in the neighbour list. This means a thread iterates through a list of unprocessed neighbours until it finds a particle that is accessible and computes the interaction for that one.\footnote{We note that we do not create copies of particles in the primary phase as done for other physics modules in \textsc{OpenGadget3} to build the neighbour list. Instead, we use references, which reduces computational costs slightly and allows us to ensure that a particle has only one velocity that gets updated and used for the subsequent interaction.}

If one implements the parallelisation as described above, one needs to be careful not to risk a deadlock. This would, for example, be the case when thread A is processing particle $i$ with particle $j$ being part of its neighbour list, and at the same time, thread B is processing particle $j$ with particle $i$ being part of its neighbour list. In this case, thread A locks particle $i$ and thread B locks particle $j$, but A cannot access particle $j$ and B cannot access particle $i$. We avoid deadlocks by always closing the lock for the particle with the larger ID first and the other particle second.

The explanations above apply to the primary phase. However, we also use the OpenMP parallelisation for the secondary phase, where the interactions for particles received from another process are computed. Their neighbour lists contain only local particles, which simplifies the problem.
We only need to use locks for the neighbour list particles, i.e.\ the local ones, but do not need locks for the received particles. Hence, we have to close only one lock per pairwise interaction.

\subsection{Conserved quantities} \label{sec:conserved_quantities}

In the SIDM module of \textsc{OpenGadget3}, the pairwise interactions are formulated in a way that each interaction conserves linear momentum and energy explicitly. This is not violated by multiple scatterings. In particular, the parallelisation of the computations preserves this property. This means that if only self-interactions are simulated and not other physical processes, such as gravity, are included, all the error in linear momentum and energy arises from round-off errors, no matter how many OpenMP threads or MPI ranks are used.

We want to note that only a few SIDM implementations have an explicit conservation of the total energy \citep[see for example][]{Valdarnini_2024}. In practice, multiple scattering leads in many codes to an artificial heating as had been pointed out by \citep{Robertson_2017a}. This problem arises when a particle scatters more than once per time step, but each scattering event is computed based on its velocity at the beginning of the time step. If the velocity is not updated between the scatterings, for example, only the velocity changes due to each scattering event are summed up, and at the end of the time step, applied onto the particle, energy is not conserved.

Despite the excellent energy and linear momentum conservation in our implementation, we have to note that angular momentum is not explicitly conserved. This is because, when the velocity kicks are applied to the particles of an interacting pair, there is a non-zero finite distance between the two interaction partners. As the error in angular momentum depends on this distance, a smaller kernel size would reduce it accordingly. Hence, we expect that the simulation results nevertheless converge.
Interestingly, \cite{Kochanek_2000} argued that the error in conservation of angular momentum per scattering event should be random and thus sum to zero over all interactions. However, whether this argument holds in general or only for specific set-ups has not been tested.

\subsection{Limitations} \label{sec:limitations}

In the following, we comment on some limitations that the numerical formulation that we employ in \textsc{OpenGadget3} faces. These should be taken into account when designing simulations.

There is a limitation on the ratio of kernel size and mean free path that can be simulated. \cite{Koda_2011} pointed out that the kernel size must be smaller than the mean free path, because otherwise particles separated by more than the mean free path could interact with each other, leading to unphysical effects. In the case of the gravothermal evolution of a DM halo, this would lead to a too large heat transfer.
This is in line with the findings by \cite{Fischer_2025b}. Following that work, we can express the ratio of kernel size to mean free path as
\begin{equation}
    \frac{h}{l} = \frac{\sigma_\mathrm{eff}}{m} \, \rho^{2/3} \, \sqrt[3]{\frac{3 \, m_\mathrm{n} \, N_\mathrm{ngb}}{\uppi \, \sqrt{2}}} \,.
\end{equation}
We use the effective cross-section, $\sigma_\mathrm{eff} = \langle \sigma_\mathrm{V,norm}(v) \, v \rangle / \langle v \rangle$, an angle- and velocity-averaged quantity \citep{Yang_2022D}. Depending on the exact problem, it is not always required that $h/l < 1$, as discussed by \cite{Fischer_2024b} and demonstrated by \cite{Fischer_2025b} for the gravothermal evolution of an isolated halo.

Another limitation may arise for cross-sections that become fairly large at a finite velocity. For example, for resonant scattering, the cross-section could be large within a small velocity range, while the overall effect of SIDM is small. However, the scatterings within that small velocity range could imply fairly small time steps and substantially slow down the simulation.

Additionally, we note that simulating a model with fairly unequal mass ratios can be challenging. The requirement that the numerical mass ratio must match the physical mass ratio can make it difficult to achieve a sufficient mass resolution for the heavier particles. Moreover, in simulations involving gravity, artificial mass segregation can arise from gravitational interactions. Mitigating this artefact may require increasing the resolution of the simulation. These problems could make the costs for a simulation with very unequal mass ratios prohibitively expensive.

Finally, we want to mention that \textsc{OpenGadget3} in general does not guarantee bitwise reproducibility.
This is, for example, true for the OpenMP parallelisation of the SIDM module.
Bitwise reproducibility may only be obtained in special cases, which can require switching off the topology-aware optimisation of the MPI computations.

\begin{figure*}
    \centering
    \includegraphics[width=\linewidth]{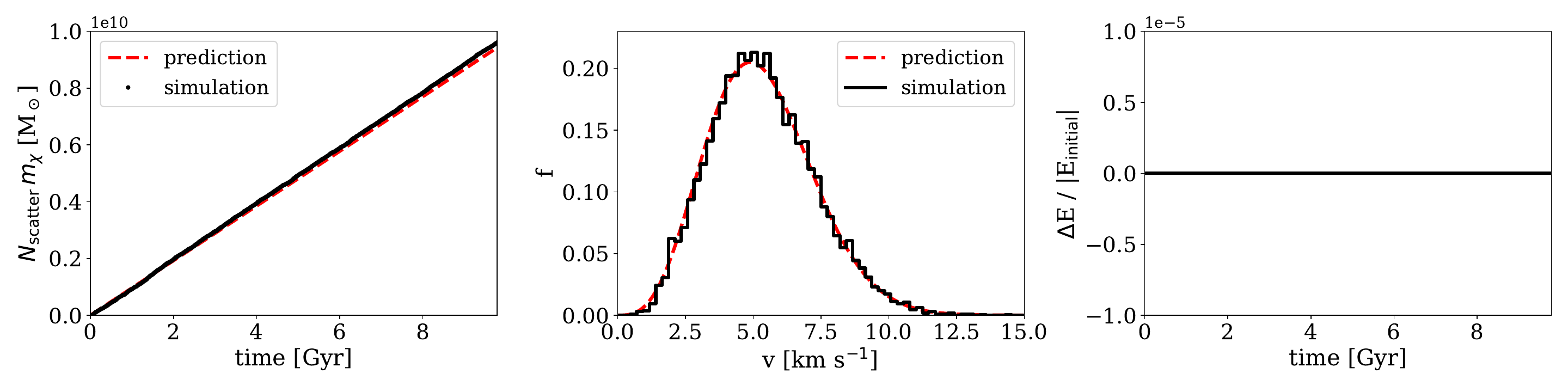}
    \caption{Test problem for rSIDM with scattered mass (left panel), distribution of scatter velocities (middle panel), and energy conservation error (right panel). The number of scatter events times the physical DM particle mass since the start of the simulation is shown as a function of time in the left panel. Here, the analytic prediction (Eq.~\eqref{eq:scatter_rate}) is indicated in red and the simulation result is in black. The middle panel gives the distribution of the relative velocities of all scattering events that occurred in the simulation. Again, we show the analytic expectation (Eq.~\eqref{eq:vel_scat_dist}) and the simulation result. The right panel shows the error in energy conservation relative to the initial energy.}
    \label{fig:test_problem_scatter_count}
\end{figure*}

\subsection{Particle physics beyond our implementation} \label{sec:physics_beyond_implementation}
\textsc{OpenGadget3} allows the simulation of a variety of SIDM models. However, more models could be implemented. This not only includes additional differential cross-sections with elastic scattering. Also, models implemented in other codes could be of interest. These include processes giving rise to inelastic scattering as previously studied by \cite{Huo_2020, Shen_2021, Xiao_2021} or models that contain long-lived excited states \citep{Vogelsberger_2019, Chua_2020, ONeil_2023, Leonard_2024, Low_2026}.


\section{Test problems} \label{sec:test_problems}

In this section, we present several tests performed with the SIDM module of \textsc{OpenGadget3}. We begin with a few tests to verify the implementation and continue with studying how the code scales for parallel computations. Lastly, we show simulations of the gravothermal evolution in the collapse regime of an isolated halo.

\subsection{Verification tests}

Several tests with known solutions have been used to check the SIDM implementation in \textsc{OpenGadget3}. They have been presented in a few publications \citep{Fischer_2021a, Fischer_2022, Fischer_2024a, Arido_2025, Patil_2025}. In the following, we show additional tests. They all have in common that they do not involve gravity but self-interactions only.

\subsubsection{Scatter count and velocities}

In this first test, we check if the number of scattering events and the distribution of the scattering velocities are correct for the rSIDM implementation using a simple test set-up.
The ICs consist of a constant density ($\rho = 10^{7} \, \mathrm{M_\odot} \, \mathrm{kpc}^{-3}$) with velocities following a Maxwell-Boltzmann distribution with a one-dimensional velocity dispersion of $\nu = 2 \, \mathrm{km} \, \mathrm{s}^{-1}$. The $10^4$ particles are inside a cube with a side length of $10 \, \mathrm{kpc}$ and periodic boundary conditions. The system is evolved without gravity, only self-interaction with an isotropic velocity-independent cross-section of $\sigma/m_\chi = 20 \, \mathrm{cm}^2 \, \mathrm{g}^{-1}$.
For SIDM, we use a spline kernel \citep{Monaghan_1985} and $N_\mathrm{ngb} = 64$.
Given that the system is already in equilibrium, it does not evolve. This allows us to easily predict the number of scattering events as well as their velocity distribution.

The number of physical scattering events since the beginning of the simulation as a function of time is
\begin{equation} \label{eq:scatter_rate}
N_\mathrm{scatter}(t) = \sqrt{\frac{4}{\uppi}} \, \frac{\sigma}{m_\chi} \, \rho \, t \, \nu \, N_\chi\,.
\end{equation}
This equation is only valid for velocity-independent and isotropic cross-sections.
Moreover, $\nu$ denotes the distribution parameter of the initial Maxwell-Boltzmann distribution, i.e.\ the one-dimensional velocity dispersion.
The total number of physical DM particles is given by $N_\chi$.

The distribution of scattering velocities for a velocity-independent cross-section is
\begin{equation} \label{eq:vel_scat_dist}
    f(v_\mathrm{scat}) = \frac{1}{8} \frac{v_\mathrm{scat}^3}{\nu^4} \exp\left(-\frac{v_\mathrm{scat}^2}{4 \, \nu^2}\right) \,.
\end{equation}
Here, $v_\mathrm{scat}$ is the relative velocity of the scattering particles. It is preserved during a single scattering event, as we assume elastic scattering.

The results for the test problem are displayed in Fig.~\ref{fig:test_problem_scatter_count}. Here, we find that the scattering rate (left panel) as well as the velocity distribution of the scattering events (middle panel) agree well with the exact solution.
Moreover, the energy is perfectly conserved (right panel).
We want to point out that it is not only important to correctly reproduce the scattering rate, but also the velocity distribution. It is known that not only does the scattering probability depend on the velocity, but also does the impact that a scattering event has on the evolution of, for example, an isolated halo. In consequence, if the distribution of scattering velocities is not correct, this could lead to a too-strong or too-weak effect of SIDM. This implies that one needs to be careful if one wants to order the neighbour list of scattering partners according to a specific criterion, such as the scattering probability, before computing the scattering events to ensure that the scheme is consistent with the physics.

\subsubsection{Angular deflection test: Single event M{\o}ller scattering} \label{sec:test_angular_dist_moeller}

In the second test problem, we consider a beam of particles that scatter in a target. We allow them to scatter only once over the course of the simulation\footnote{To realise this, we have implemented an additional condition that checks if a particle has scattered at an earlier time before we would allow it to scatter.} and compare the distribution of their deflection angles to the differential cross-section. This allows us to check whether the scattering angle is sampled correctly.

For the ICs, we choose a setup with $8\,000$ beam particles that scatter in a target consisting of $92\,000$ particles. The target is a cube of side length $14 \, \mathrm{kpc}$ with a DM density of $\rho = 3.353 \times 10^6 \, \mathrm{M_\odot} \, \mathrm{kpc}^{-3}$. The beam particles have an initial velocity of $v_\mathrm{ini} = 2 \, \mathrm{km} \, \mathrm{s}^{-1}$.
These are the same ICs as used in sect.~3.3 by \cite{Fischer_2021a}.

\begin{figure}
    \centering
    \includegraphics[width=\columnwidth]{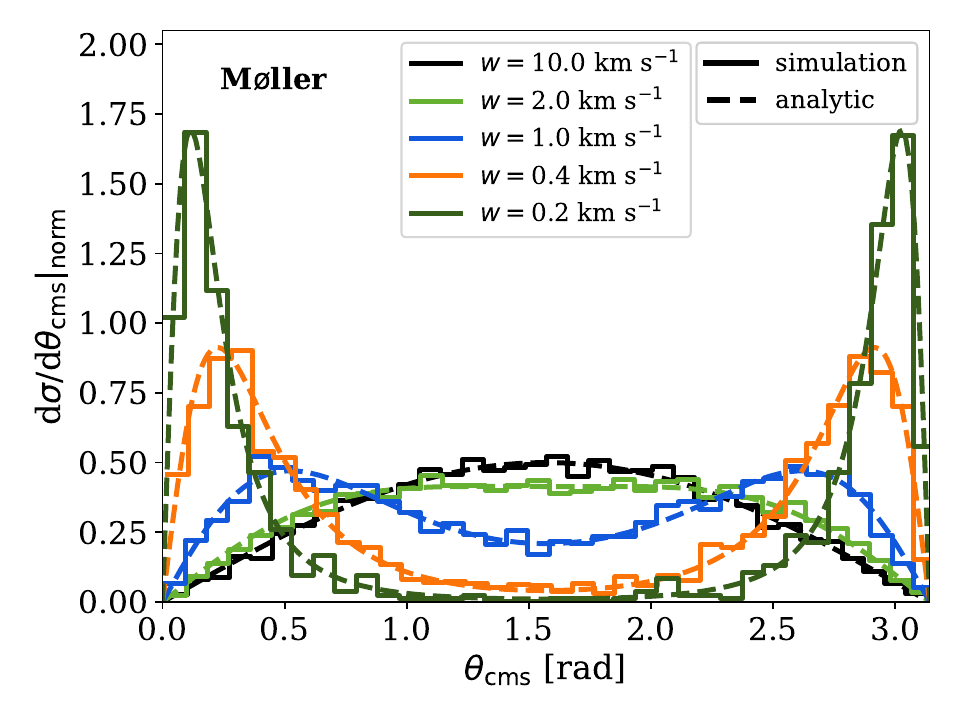}
    \caption{Distribution of deflection angles for M{\o}ller scattering. The distribution of the scattering angles as a function of the scattering angle is shown and compared to the analytic expression of the M{\o}ller cross-section. Solid lines are for the simulation results and dashed lines are for the exact solution given by Eq.~\eqref{eq:moeller_dcs}. The colours refer to different values for the parameter $w$, which we have tested.}
    \label{fig:angular_dist_moeller}
\end{figure}

We simulate the problem using M{\o}ller scattering (Eq.~\eqref{eq:moeller_dcs}) with different parameters for $w$. The results are shown in Fig.~\ref{fig:angular_dist_moeller}.
Here, we compare the simulation results to the exact solution. Overall, the simulations match the predictions well. Hence, the SIDM module in \textsc{OpenGadget3} properly samples the scattering angles. 

\subsubsection{Angular deflection test: Fixed target Rutherford scattering} \label{sec:test_angular_dist_rutherford}

The third test problem considers a beam with particles scattering of a target as in Sect.~\ref{sec:test_angular_dist_moeller}. But now we allow the particles to scatter multiple times. Importantly, we assume a fixed target, i.e.\ the target particles are assumed to be infinitely more massive than the beam particles. For this, we modify the code as described by \cite{Arido_2025}. 

\begin{figure}
    \centering
    \includegraphics[width=\columnwidth]{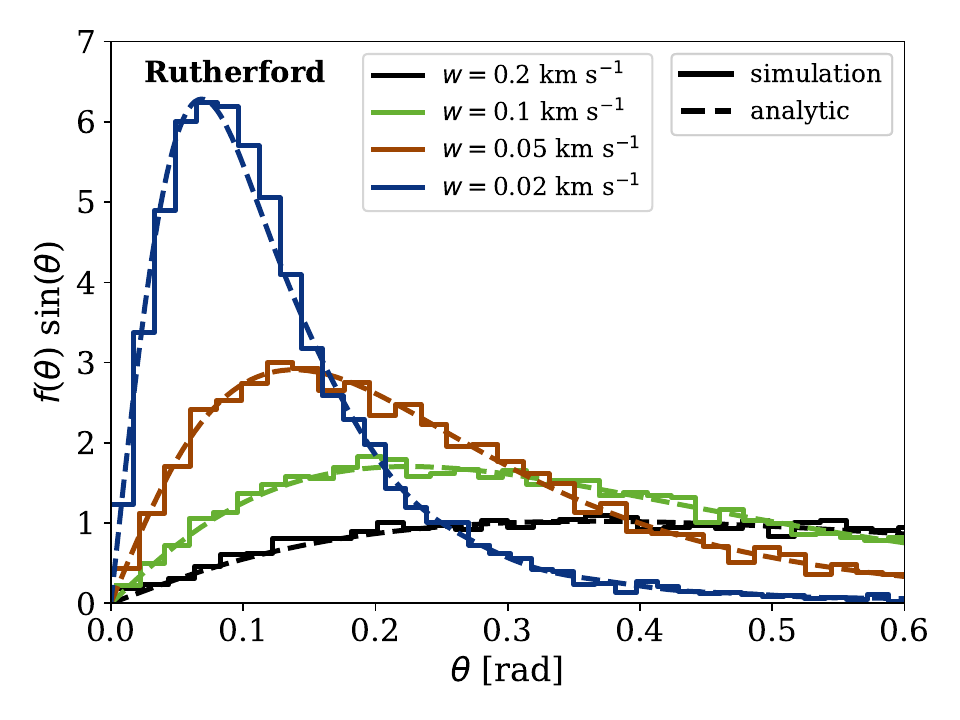}
    \caption{Distribution of deflection angles for Rutherford scattering. We simulated Rutherford scattering employing different parameters and compare them to the exact solution from Eq.~\eqref{eq:goudsmit+saunderson}. The velocity parameter $w$ is specified in the legend, and $\sigma_0 / m$ times the time for which we show the results is
    $t \, \sigma_0 / m = \{3.92 \times 10^5, 1.96 \times 10^6, 9.85 \times 10^6, 7.84 \times 10^7\} \, \mathrm{cm}^2 \, \mathrm{g}^{-1} \, \mathrm{Gyr}$ (black, light green, brown, dark blue).}
    \label{fig:angular_dist_rutherford}
\end{figure}

We use the same ICs as for the previous test with M{\o}ller scattering (Sect.~\ref{sec:test_angular_dist_moeller}). This time we simulate Rutherford scattering (Eq.~\eqref{eq:rutherford_dcs}) choosing different values for $\sigma_0$ and $w$. 
The results are shown in Fig.~\ref{fig:angular_dist_rutherford}. Here, we compare the distribution of the simulated deflection angles to the expected distribution. The latter can be computed according to \cite{Goudsmit_1940a, Goudsmit_1940b}. Following the derivation by \cite{Bethe_1953} one finds
\begin{align} \label{eq:goudsmit+saunderson}
    f(t,\theta) = \sum^\infty_{l=0} \,&P_l(\cos{\theta})\left(l+\frac{1}{2}\right) \exp\bigg\{-n \, v \, t \nonumber \\
    & \int \frac{\mathrm{d}\sigma}{\mathrm{d}\cos\theta'} \left[ 1 - P_l(\cos{\theta'}) \right] \, \mathrm{d}\cos\theta' \bigg\} \, ,
\end{align}
where $P_l(x)$ are Legendre polynomials and $n$ is the number density of the target particles.

The simulations were run with the hybrid scheme employing $\theta_\mathrm{c} = 0.1$ (see Sect.~\ref{sec:differential_cross_section}). This is especially helpful in speeding up the computations for fairly anisotropic scattering. In general, we find in Fig.~\ref{fig:angular_dist_rutherford} that the simulations agree well with the predictions.

\subsubsection{Comoving integration test}

\begin{figure}
    \centering
    \includegraphics[width=\columnwidth]{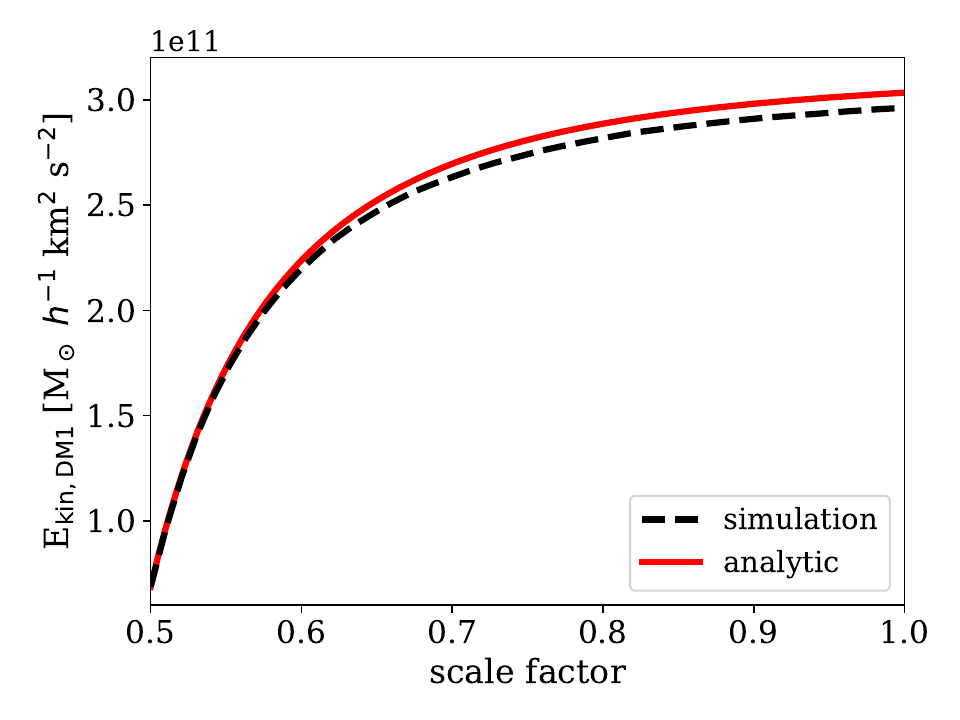}
    \caption{Heat conduction problem for comoving integration. The kinetic energy of the initially cooler DM component in terms of the canonical momentum is shown as a function of the scale factor. The simulation results are given by the black dashed line and compared to the exact solution (Eq.~\eqref{eq:heat_conduction_cosmo}) shown in red.}
    \label{fig:comoving_integration_test}
\end{figure}

As a last test for which we know the solution, we consider two components in an expanding space and compute the heat exchange between them. The two components initially follow the same constant density; each component consists of $10^5$ particles. The velocities of the two components are drawn from Maxwell-Boltzmann distributions, where one has a higher temperature than the other. As we assume equal mass scattering ($r=1$), we expect that the two components exchange heat and approach the same temperature. The initial energy of the colder DM component is $E_\mathrm{DM1} = 2.75 \times 10^{11} \, \mathrm{M_\odot} \, h^{-1} \, \mathrm{km}^2 \, \mathrm{s}^{-2}$, and the energy of the warmer component is $E_\mathrm{DM2} = 2.75 \times 10^{12} \, \mathrm{M_\odot} \, h^{-1} \, \mathrm{km}^2 \, \mathrm{s}^{-2}$.
We note that this is the same test problem as considered by \cite{Fischer_2025a} in their appendix~D, but now for DM self-interactions instead of DM-baryon scattering.

We simulate the test problem with a forward-dominated and velocity-independent cross-section of $\sigma_\mathrm{T} / m_\chi = 2 \times 10^6 \, \mathrm{cm}^2 \, \mathrm{g}^{-1}$.
The cross-section is chosen to have a larger effect and thus is much larger than what current constraints suggest.
In Fig.~\ref{fig:comoving_integration_test}, the simulation results are shown and compared with the exact solution. In particular, we show the time evolution of the kinetic energy of the initially colder DM component.
It is visible that the simulation can reproduce the expected behaviour, especially at small scale factors it agrees very well.
We note that the analytic solution makes the assumption that the velocities of each component always follow a Maxwell-Boltzmann distribution, which is only an approximation and strictly speaking not fulfilled.

We obtained the exact solution by building on the work by \cite{Dvorkin_2014}. In the non-expanding case, the power $P_\mathrm{DM}$ with which the energy of one of the two DM components changes over time is given by
\begin{align} \label{eq:heat_conduction_static}
    P_\mathrm{DM1} =& \frac{\mathrm{d}E_\mathrm{DM1}}{\mathrm{d}t} \nonumber\\
    =& -\frac{32}{\sqrt{27\uppi}} \, \frac{\rho_\mathrm{DM1} \, \rho_\mathrm{DM2}}{(1+r)^2} \frac{\sigma_\mathrm{T}}{m_\chi} \left(\frac{E_\mathrm{DM1}}{M_\mathrm{DM1}} + \frac{E_\mathrm{DM2}}{M_\mathrm{DM2}}\right)^{1/2} \nonumber \\
    &\times \left[ E_\mathrm{DM1} \left( \frac{1}{\rho_\mathrm{DM1}} + \frac{r}{\rho_\mathrm{DM2}} \right) - \frac{r E_\mathrm{tot}}{\rho_\mathrm{DM2}} \right] \,.
\end{align}
In the case of an expanding space, as in our comoving integration test, the differential equation describing the evolution of the system is
\begin{equation} \label{eq:heat_conduction_cosmo}
    \frac{\mathrm{d}E_\mathrm{DM1}}{\mathrm{d}a} = \frac{P_\mathrm{DM1}(a)}{a \, H(a)} -\frac{2 E_\mathrm{DM1}}{a} \,.
\end{equation}
Here, $a$ denotes the scale factor and $H(a)$ is the Hubble parameter.
Moreover, $P_\mathrm{DM1}(a)$ is given by Eq.~\ref{eq:heat_conduction_static}. 
The quantities in Eq.~\eqref{eq:heat_conduction_static} and~\eqref{eq:heat_conduction_cosmo} are in terms of their peculiar values, i.e.\ $\rho \propto a^{-3}$ and $E \propto a^{-2}$.
We obtained the solution shown in Fig.~\ref{fig:comoving_integration_test} by numerically integrating Eq.~\eqref{eq:heat_conduction_cosmo}.

\subsection{Scaling tests} \label{sec:scaling_tests}

In this section, we test the scaling of the SIDM implementation by performing strong scaling tests. First, we simulate an isolated halo testing the OpenMP and MPI parallelisation (Sect.~\ref{sec:scaling_tests_halo}). Second, we perform a strong scaling test for a full box cosmological simulation (Sect.~\ref{sec:scaling_tests_cosmo}).

\subsubsection{Isolated halo} \label{sec:scaling_tests_halo}

For the scaling test with an isolated halo, we use ICs following a Navarro--Frenk--White \citep[NFW][]{Navarro_1996} profile. It is given by
\begin{equation}
    \rho(r) = \frac{\rho_0}{\frac{r}{r_\mathrm{s}} \left( 1 + \frac{r}{r_\mathrm{s}} \right)^2} \,.
\end{equation}
For the density parameter, we chose $\rho_0 = 4.42 \times 10^7 \, \mathrm{M}_\odot \, \mathrm{kpc}^{-3}$ and set the scale radius to $r_\mathrm{s} = 1.28 \, \mathrm{kpc}$. Additionally, the halo is truncated at $r_\mathrm{cut} = 15 \, r_\mathrm{s}$. The ICs, containing $N = 10^7$ particles, were generated using \textsc{SpherIC} \citep{Garrison_Kimmel_2013} and have been used before in \cite{Fischer_2025b}.
For the simulations, we set the gravitational softening length to $\epsilon = 3.0 \, \mathrm{pc}$, use the $N_\mathrm{ngb} = 48$ nearest neighbours for the SIDM computations, and choose isotropic velocity-independent scattering with a total cross-section per DM particle mass of $\sigma / m = 30 \, \mathrm{cm}^2 \, \mathrm{g}^{-1}$.

\begin{figure}
    \centering
    \includegraphics[width=\columnwidth]{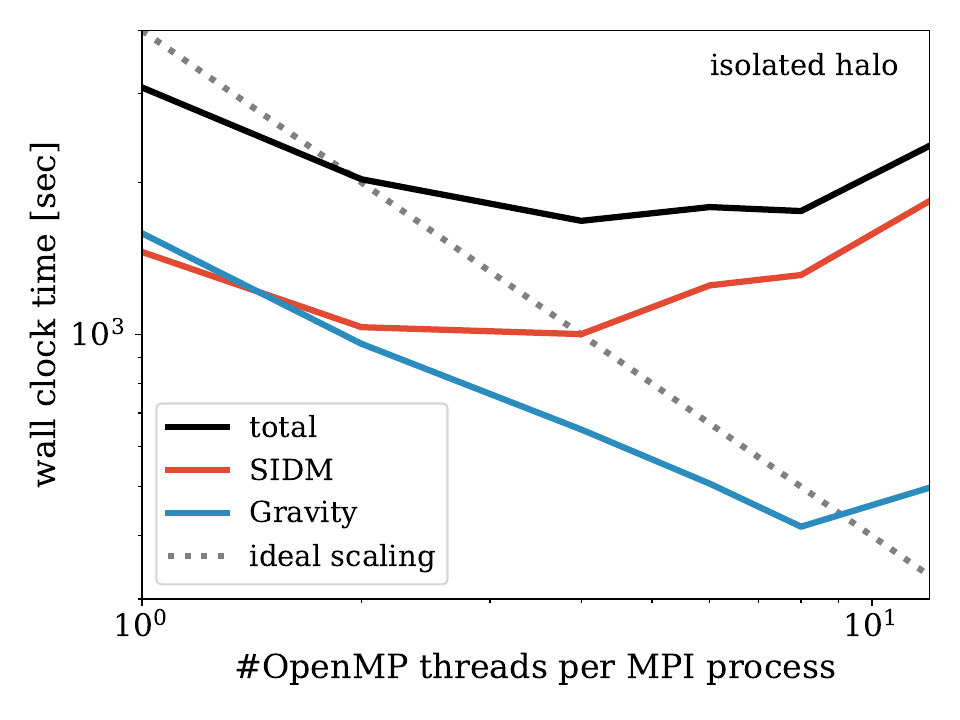}
    \caption{Strong scaling test for the OpenMP parallelisation. We show the wall clock time as a function of the number of OpenMP threads, while the number of MPI processes is fixed to 18. The total wall clock time to advance the simulation by almost $10 \, \mathrm{Myr}$ is shown in black, whereas the time required for the SIDM and gravity computations is shown in red and blue, respectively. The gray dotted line indicates a scaling proportional to the inverse of the number of threads.}
    \label{fig:halo_openmp}
\end{figure}

\begin{figure}
    \centering
    \includegraphics[width=\columnwidth]{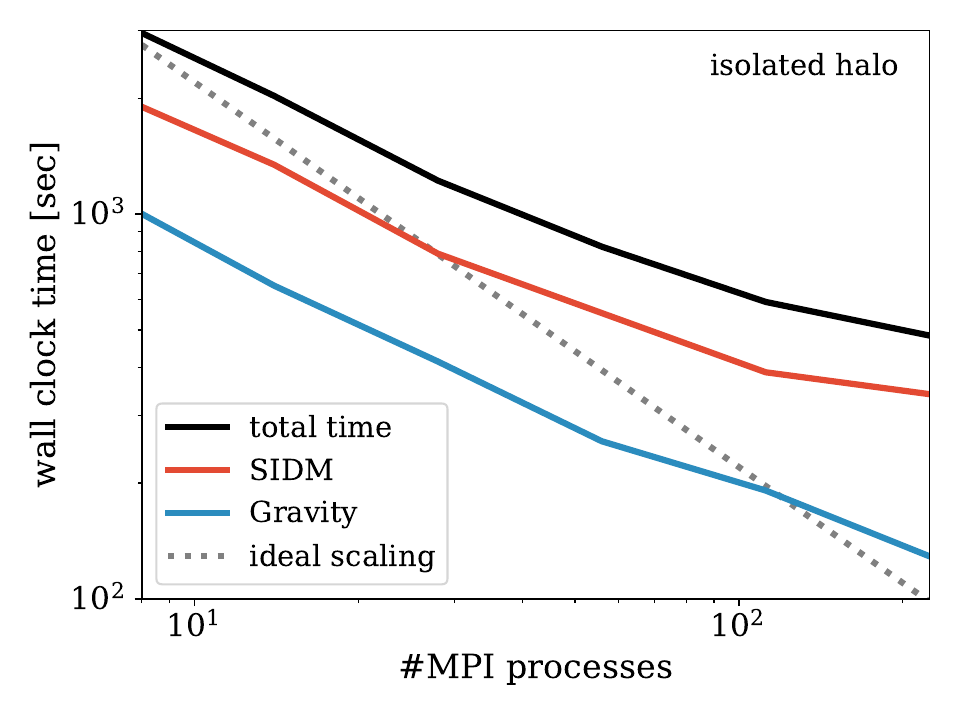}
    \caption{Strong scaling test for the MPI parallelisation. Similar to Fig.~\ref{fig:halo_openmp}, we show the wall clock time required to advance the simulation by almost $10 \, \mathrm{Myr}$, but this time as a function of the number of MPI processes. At the same time, we keep the number of OpenMP threads fixed to four.}
    \label{fig:halo_mpi}
\end{figure}

We begin with testing the scaling of the OpenMP parallelisation. Therefore, we simulate the halo for almost $10 \, \mathrm{Myr}$ using 18 MPI processes and vary the number of OpenMP threads.
In Fig.~\ref{fig:halo_openmp}, we give the results distinguishing the computing time spent on gravity and SIDM. It is well visible that the OpenMP parallelisation does not scale well with the number of threads, especially for a larger number of threads, it can slow down the computations. Nevertheless, it allows one to get a substantial speed-up when a few threads are used. For example, using four OpenMP threads and leaving the rest of the parallelisation to MPI can be a reasonable choice.
This can be especially useful if one is limited in memory and cannot afford the additional memory required to use more MPI ranks or when one wants to keep the computational costs for parts of the code low that do not scale very well with the number of MPI ranks. For example, the costs for the domain decomposition scale quadratically with the number of MPI ranks.
Following this, we test the MPI parallelisation next.

We keep the number of OpenMP threads fixed to four and vary the number of MPI processes only. Using the same test problem as before, we obtain the results shown in Fig.~\ref{fig:halo_mpi}. We find that the scaling becomes worse as the number of MPI processes increases. Here, the workload per process becomes less compared to the parallelisation overhead. Moreover, it is visible that the SIDM computations require about twice the computing time that gravity needs.

\subsubsection{Cosmological simulation} \label{sec:scaling_tests_cosmo}

Here, we consider a full cosmological box to perform a strong scaling test of the MPI parallelisation. The ICs are generated with \textsc{N-GenIC} \citep{Springel_2015} and consist of $N = 576^3$ particles in a cubic box with a side length of $l_\mathrm{box} = 48 \, \mathrm{cMpc} / h$. The following cosmological parameters are employed: $\Omega_\mathrm{M} = 0.272$, $\Omega_\mathrm{\Lambda} = 0.728$, $h = 0.704$, $n_\mathrm{s} = 0.963$, and $\sigma_8 = 0.809$ \citep[WMAP7,][]{Komatsu_2011}. The initial conditions consist of DM particles only, and all of them have the same mass. We simulated isotropic and velocity-independent scattering with a total cross-section of $\sigma / m = 1 \, \mathrm{cm}^2 \, \mathrm{g}^{-1}$ using $N_\mathrm{ngb} = 48$ nearest neighbours.

\begin{figure}
    \centering
    \includegraphics[width=\columnwidth]{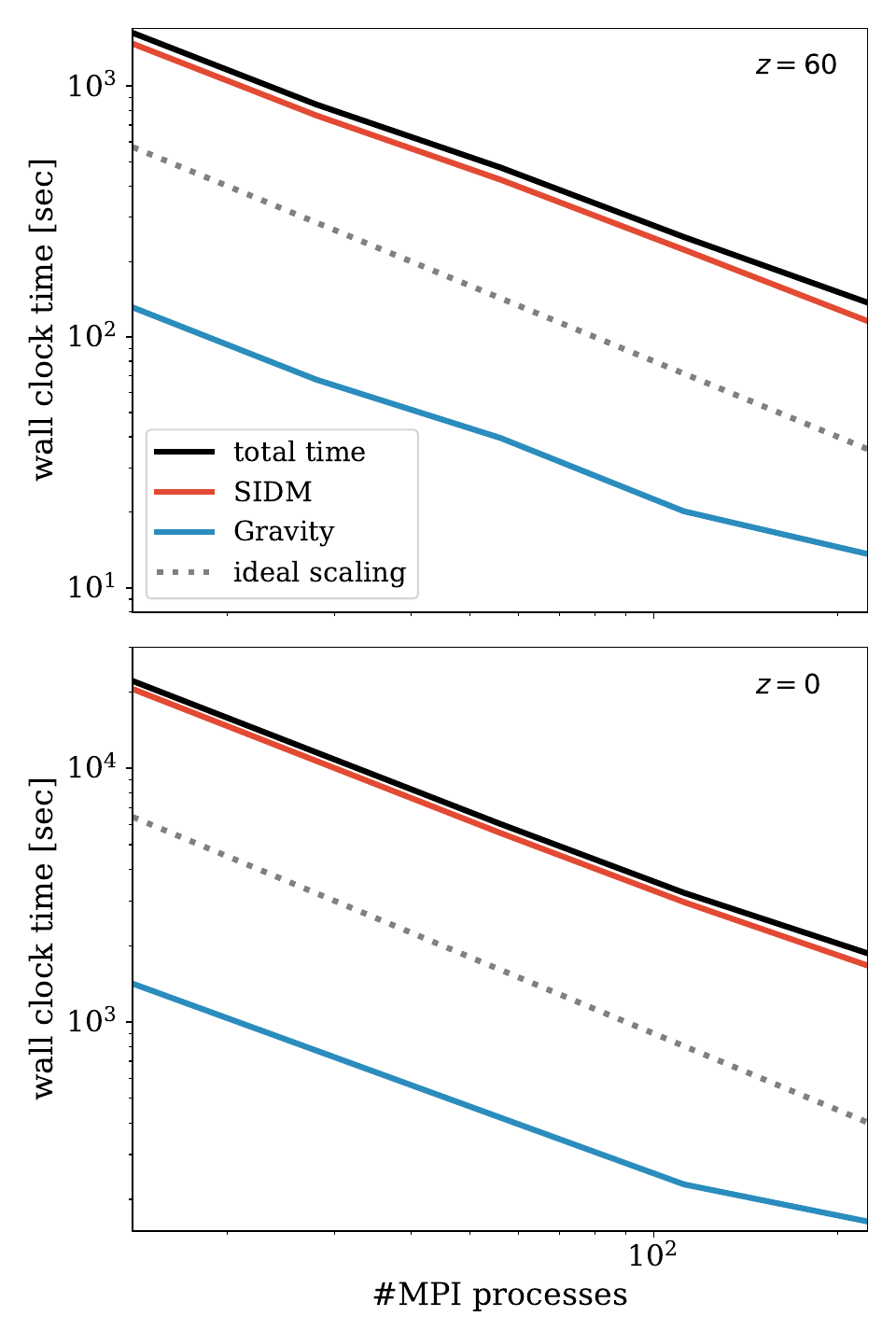}
    \caption{Strong scaling test for the MPI parallelisation. We simulate a full cosmological box as described in the main text using four OpenMP threads and varying the number of MPI processes. The wall clock time needed to advance the simulation for a few steps is shown when starting from a redshift of $z = 60$ (upper panel) and $z=0$ (lower panel). We indicate the total time spent in black, and the SIDM and gravity-related costs in red and blue, respectively.}
    \label{fig:cosmo_mpi}
\end{figure}

We display the strong scaling test in Fig.~\ref{fig:cosmo_mpi}. Here, we advanced the simulation from a redshift of $z=60$ (upper panel) and $z=0$ (lower panel) by a few time steps and measured the required wall clock time when varying the number of MPI processes. The number of OpenMP threads is fixed to four. It is visible that the scaling of the SIDM and gravity computations is close to the ideal scaling, but a little worse. Given that the SIDM computations scale well with the number of processors in the tested range, our MPI implementation is well-suited for parallel computations.

Finally, we note that the SIDM computations are more expensive relative to the gravity ones for the cosmological box compared to the isolated halo. Not only is the problem studied here different, but also the gravitational forces were computed differently. While for the isolated halo the gravitational forces were computed based solely on an oct-tree, it was only used for short-range forces in the cosmological simulations, whereas the long-range forces were evaluated based on a particle mesh \citep[for a detailed description see][]{Dolag_2026}. This may contribute to the difference that we see here.
In addition, when considering full-physics simulations, the costs for the baryonic physics outweigh the SIDM costs.

\subsection{Gravothermal collapse} \label{sec:collapse}
A challenging problem in modelling SIDM is posed by the gravothermal collapse of DM halos. In a gravitationally self-bound halo, the self-interactions lead to energy transport following the velocity dispersion gradient. This implies an outward energy flux at sufficiently larger radii. The local thermalisation of the velocity distribution due to the scatterings can even produce particles with speeds above the escape velocity, i.e.\ they become unbound and leave the system. Due to the loss of energy, the system collapses, reaching a very high density and velocity dispersion in its centre.

This gravothermal collapse could provide an interesting mechanism for forming highly concentrated objects that could explain several observations \citep[e.g.][]{Yu_2026}, including perturbations by substructure in gravitational strong lenses. However, it has been found to be challenging to simulate \citep[e.g.][]{Yang_2022D, Zhong_2023, Fischer_2024b, Mace_2024, Palubski_2024}.

\begin{figure}
    \centering
    \includegraphics[width=\columnwidth]{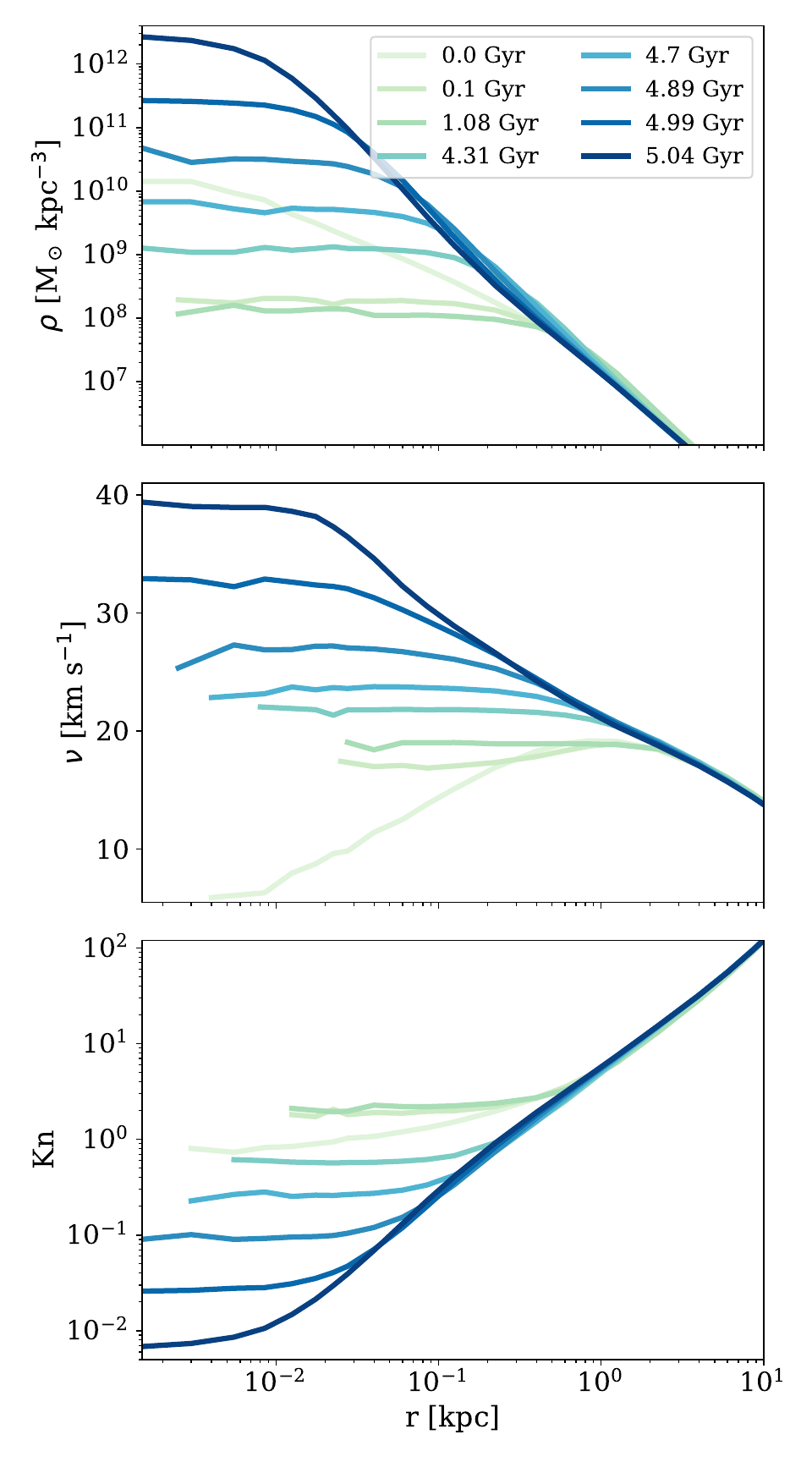}
    \caption{Gravothermal evolution of an isolated halo. Various quantities are shown as a function of radius for a halo evolving under the influence of DM self-interactions. The top panel gives the density profile, the middle one displays the one-dimensional velocity dispersion, and the bottom one shows the Knudsen number.}
    \label{fig:profiles}
\end{figure}

\begin{figure}
    \centering
    \includegraphics[width=\columnwidth]{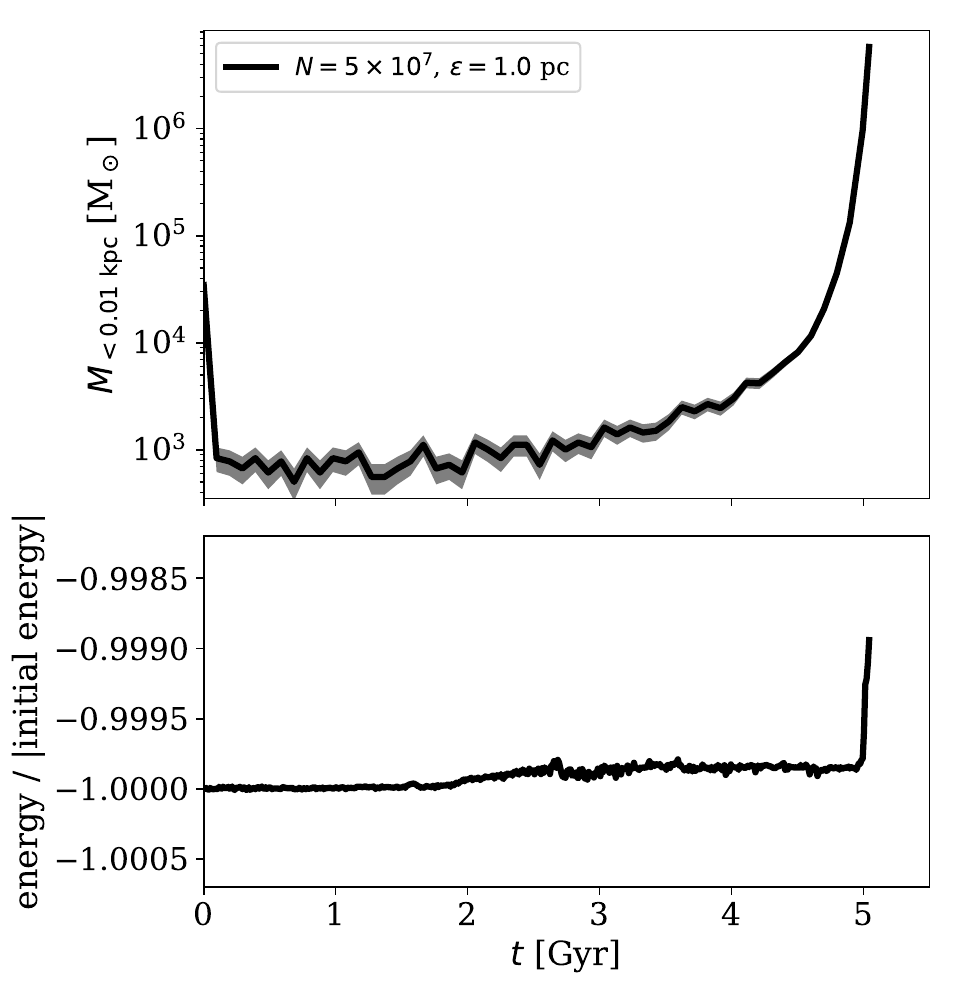}
    \caption{Enclosed mass and energy conservation for an isolated SIDM halo. The upper panel gives the mass enclosed within $10 \, \mathrm{pc}$ as a function of time. The lower panel displays the energy conservation, namely the ratio of the total energy to the absolute of its initial value as a function of time.}
    \label{fig:encmass}
\end{figure}

A high-resolution simulation ($N=5\times10^7$) of a collapsing DM halo with \textsc{OpenGadget3} had been publicly released by \cite{Fischer_2025b}. This simulation can serve as a benchmark for other codes and methods.\footnote{The simulation data can be found at \url{https://darkium.org}.}
The ICs have the same properties as the ones described in Sect.~\ref{sec:scaling_tests_halo}, but are resolved with $N=5\times 10^7$ particles. The simulation is run with a gravitational softening length of $\epsilon = 1 \, \mathrm{pc}$ and a velocity-independent isotropic scattering with a total cross-section of $\sigma / m = 80 \, \mathrm{cm}^2 \, \mathrm{g}^{-1}$. 
Such a large cross-section is clearly ruled out at large velocities ($v \approx 1000 \, \mathrm{km} \, \mathrm{s}^{-1}$) given current galaxy cluster constraints \citep[e.g.][]{Andrade_2021, Sagunski_2021, Eckert_2022, Gopika_2023, Adhikari_2025a, ODonnell_2026}, but is reasonable at lower velocities encountered in low mass halos such as ours with a mass of $\approx 2.8 \times 10^9 \, \mathrm{M_\odot}$ \citep[e.g.][]{Gilman_2021, Silverman_2022, Zhang_2024}. Hence, our cross-section can be interpreted as an effective cross-section representative of a realistic cross-section with a velocity dependence fulfilling the current constraints.

Here we continued the simulation for an additional time of $0.05\,\mathrm{Gyr}$ and show the density, velocity dispersion, and Knudsen number for several times in Fig.~\ref{fig:profiles}.
It can be seen in the upper panel that the density in the centre of the halo drops at the beginning. This can be understood by heat flowing into the central region caused by the positive velocity dispersion gradient, as visible in the middle panel.
Later, the velocity dispersion gradient becomes flat or negative at all radii, and heat can only flow outward; the density core of the halo is collapsing.
Along with the increase of density and velocity dispersion, the mean free path of the particles shrinks relative to the gravitational scale height. The Knudsen number, given by the ratio of the two, becomes fairly small as shown in the lower panel.

In addition, we display the evolution of the mass enclosed within $10 \, \mathrm{pc}$ in the upper panel of Fig.~\ref{fig:encmass}. Here, the core formation with its decrease of the enclosed mass can be seen as well as the collapse phase with an increase in the enclosed mass.
The lower panel shows the energy conservation. The total energy is well conserved until the end of the simulation.
Here, our parallelisation scheme (see Sect.~\ref{sec:parallelisation}) is very helpful as it allows us to deal with particles scattering multiple times per time step and does not give rise to an artificial heating term due to energy non-conservation from multiple scatterings \citep{Robertson_2017a}.
However, we can clearly see that the error in total energy drastically increases in the late collapse phase, although this is a very high-resolution simulation.
The reasons for energy non-conservation have been discussed in detail by \cite{Fischer_2024b}.
We note that simulating deep into the collapse phase is not only limited by the energy conservation, but would also require a better spatial resolution to probe deeper into the gravothermal catastrophe. This limitation arises from resolving the mean free path as discussed in Sect.~\ref{sec:limitations}.
In conclusion, we can accurately simulate the collapse of SIDM halos; the deeper we want to simulate into the collapse, the more expensive it becomes.

\section{Discussion} \label{sec:discussion}

In this section, we discuss the properties of the numerical formulation of our SIDM scheme and challenges related to it. Moreover, we point out potential future directions for the development of SIDM simulation codes, including ways that could help to speed up the computations. 

\subsection{Numerical properties and challenges}

An interesting aspect of the numerical scheme that we have presented is the following. The interaction probability (in the rare scheme) and the drag force (in the frequent scheme) depends on the relative velocity. This relative velocity does not change due to the interaction of a single pair, but remains conserved. Hence, the scattering probability or drag force computed based on the post-scattered velocities is the same as the one based on the pre-scattered velocities. This contributes to the stability of the numerical scheme but is true only for elastic scattering. Models featuring inelastic interactions have been implemented by a loss of kinetic energy, i.e.\ reducing the relative velocity of the considered particle pair \citep[e.g.][]{Huo_2020, Shen_2021}. Simulations of such models might be more susceptible to numerical errors that, in detail, depend on the velocity-dependence of the dissipative process \citep[see Appendix~C by][]{Fischer_2024a}.

The more physically motivated simulations that we have presented here involve SIDM as well as gravity. To understand the numerical properties of such simulations, one needs to consider the interplay of the numerical schemes for gravity and self-interactions. As we mentioned in Sect.~\ref{sec:time_integration}, the gravitational accelerations in \textsc{OpenGadget3} depend only on the positions, and thus the velocity kicks from the self-interactions do not have a direct impact on them as they only alter the velocities but not the positions. This means it would not make a difference if we first compute the SIDM velocity kicks before or after gravitational accelerations. The latter would always be the same. However, this does not apply to all schemes. When the gravitational accelerations do not only depend on the positions of the particles but also on their velocities, the situation becomes more complicated, and the interplay of the gravity and self-interaction scheme could have unwanted effects. Hence, care must be taken when employing schemes such as the Hermite 4th order integrator \citep{Makino_1992}.

We want to point out that in the scheme for frequent self-interactions, many small velocity kicks are applied. Here, we found that round-off errors can lead to a substantial error overall, depending on the number precision employed. As a consequence, we recommend using double precision for the velocities of the simulation particles.

In \textsc{OpenGadget3}, we employ an adaptive time stepping scheme that allows us to evolve particles on different time steps and adapt them over the course of the simulation, as also done in many other simulation codes. We want to point out that the modelling of the self-interactions can imply a limitation on the scheme that can be used for adaptive time stepping. \cite{Springel_2021} refer to the scheme that we employ as traditional nested time integration, while they have implemented in \textsc{Gadget-4} a hierarchical time integration scheme similar to \citep{Pelupessy_2012}, especially suited for problems with a very deep time step hierarchy. Their scheme partially undoes a velocity kick due to gravity previously applied to a particle (see their eq.~(50)). This is meant to undo the gravitational acceleration arising from a subset of particles.
Partially undoing a velocity kick in SIDM can be problematic, especially if multiple interactions per particle per time step are allowed, as there is no defined SIDM acceleration as we pointed out at the end of Sect.~\ref{sec:rare_scatter}. As a consequence, the hierarchical time stepping scheme as used in \textsc{Gadget-4} or \textsc{AREPO} is not directly compatible with the SIDM scheme that we have described.
However, a little different formulation as expressed by eq.~49 in \citep{Springel_2021} could work for the SIDM computations. The more difficult problem might be the computation of the kernel size when using hierarchical time integration, which was not resolved in \cite{Springel_2021} for the SPH implementation (see their sect.~5.7).\footnote{In line with this, the SIDM simulations of the AIDA-TNG project \citep{Despali_2025} have been run with the traditional nested time integration scheme and not the hierarchical time integration.}

Robustly estimating the needed time step for the SIDM computations requires knowing the local velocity distribution. In our scheme, we gain this knowledge through all considered pairs of particles. In consequence, this can place a lower limit on the range of neighbouring particles that should be used for the SIDM computations. At the same time, a smaller neighbour number also implies smaller time steps as the value of the kernel overlap integral grows (see Eqs.~\eqref{eq:time_step_crit_rare} and~ \eqref{eq:time_step_crit_freq}). In addition, the robustness of the estimated time step can be affected by the stage at which the kernel size is computed, as we have explained in Sect.~\ref{sec:implementation_overview}.

In Sect.~\ref{sec:conserved_quantities}, we mentioned that angular momentum is not explicitly conserved. This arises from the non-zero distance between the numerical particles when they interact. Hence, the angular momentum error is expected to decrease with increasing resolution. Until now, there has been no thorough investigation of how problematic the angular non-conservation is. But the picture that it always cancels out on average might be too simple. It would be interesting to understand how large the error arising from SIDM is compared to the one already present in a collisionless simulation.
In principle, there could be several strategies to reduce the angular momentum error arising from the modelling of the self-interactions. This includes strategies to minimise the distance of the interacting particles and a choice for the azimuthal scattering angle $\varphi_\mathrm{cms}$ that is no longer random. To understand how well such strategies could work, dedicated studies are needed.

\subsection{Further directions for SIDM simulations}

After discussing numerical properties and the resulting limitations, we will turn to strategies to improve SIDM simulations. This concerns making them faster and allowing the simulation of systems that can hardly be simulated with the state of the art.

In a regime where one can safely assume a particle not to interact more than once per time step, it is not necessary to go through all pairs of neighbouring particles and the computations can be sped up by considering only a subset. This has, for example, been done in the context of a meshless DSMC code \citep[see sect.~2.3.2 by][]{Olson_2008}.

Another idea to speed up the simulations applies to the regime where the gravitational time step is significantly smaller than the SIDM time step. As we have seen in Sect.~\ref{sec:scaling_tests}, the SIDM computations are more expensive than the ones for gravity. If the SIDM computations are not applied every time step for which the gravitational accelerations are computed, but for example only every second time, this could have the chance to make the simulations much faster. The idea of decoupling the time stepping for gravity and SIDM could also be applied in the opposite regime when the SIDM time step is smaller. But here it would lead to a smaller speed-up since the gravity costs are lower compared to SIDM.

Graphics processing units (GPUs) have proven to be very powerful compared to central processing units (CPUs) for specific tasks. It has been shown that GPUs can be successfully used to accelerate collisionless $N$-body simulations involving adaptive time stepping \citep[e.g.][]{Miki_2017, Racz_2019, Ragagnin_2020}. So far, the use of GPUs in the context of SIDM simulations has not been explored. Unfortunately, the scheme that we presented here is anything but well-suited for the computations on GPUs. The same is true for other SIDM $N$-body codes. Thus, it is not a surprise that SIDM simulations have so far relied on CPUs only, and it remains to be investigated how GPUs can be utilised in this context. Interestingly, the use of GPUs has been explored in the context of DSMC codes \citep[e.g.][]{Su_2012, Goldsworthy_2014, Kashkovsky_2014, Celoria_2023}.

Another strategy to speed up SIDM simulations is to limit the scope of what is actually simulated and combine the simulation code with analytic or semi-analytic approximations. This has been explored by \cite{Zeng_2022, Zeng_2025} in the context of satellite halos. Simulating the evolution of a satellite halo when at the same time a much more massive host must be resolved is very costly. Instead, the host can be described analytically, and only the satellite halo is resolved explicitly. \cite{Zeng_2022} implemented an interaction term that describes the self-interactions between the satellite particles and the analytic host halo. Their scheme allows for efficiently simulating low-mass satellites that could hardly be studied otherwise. While this makes their approach promising, it could be extended to further particle models and employ an improved description of the host halo \citep{Klemmer_2026}.

Simulating the gravothermal collapse phase of SIDM halos is challenging due to computational costs, even though accurate simulations are possible \citep{Fischer_2025b}.
It is not only the time steps that become small in the centre of the halo, but also the length scales that need to be resolved become tiny. As discussed in Sect.~\ref{sec:limitations}, the kernel size should be small compared to the mean free path. When the mean free path is sufficiently small, SIDM starts to behave as a conducting fluid. While this could bear the chance to derive an alternative description for the central region of an SIDM halo deep in the collapse phase, it is less obvious that it could provide an adequate solution for cosmological simulations. In those, the gravothermal collapse could lead to the formation of black holes with masses that easily lie below the numerical particle mass used for DM.
In practice, those simulations fail long before a black hole may form to resolve the relevant time and length scales, which gives rise to substantial numerical errors. Eventually, building a subgrid model for the inner region of an SIDM halo could be a more practical solution for those simulations.

Finally, an interesting direction is to extend the scope beyond elastic two-to-two interactions among DM particles. This is relevant for microscopic models for DM that feature for example dissipative processes and excited states (see Sect.~\ref{sec:physics_beyond_implementation}), long-range DM self-interactions \citep[e.g.][]{DeRocco_2025}, interaction of DM with a dark radiation component, and a further exploration of multi-component scenarios with several species with different masses, allowing for example for inelastic conversion processes \citep[e.g.][]{Boddy_2014}.

\section{Conclusion} \label{sec:conclusion}

In this work, we have reviewed the SIDM implementation in the cosmological hydrodynamical simulation code \textsc{OpenGadget3} and introduced new features that have not been used before. We have described many aspects of the numerical formulation, pointed out different numerical choices across the simulation codes used for SIDM, and discussed related challenges.
In addition, we have demonstrated through several test problems that the code can accurately model DM self-interactions and is well suited for parallel computations.

Overall, the SIDM implementation in \textsc{OpenGadget3} is capable of simulating a larger variety of SIDM models, in idealised and cosmological set-ups.
The code stands out among other SIDM implementations due to several features:
\begin{itemize}
    \item its ability to simulate full differential cross-sections with their velocity and angle dependence,
    \item the inclusion of extremely forward-dominated cross-sections,
    \item its ability to simulate particle models with unequal mass ratios,
    \item a strict formulation of the SIDM time step criteria,
    \item its ability to handle multiple interactions per particle and time step,
    \item the perfect energy conservation for SIDM computations, owing to a unique parallelisation scheme.
\end{itemize}
Especially the last two points have proven to be very powerful when simulating the gravothermal collapse of SIDM halos.

With this paper, we release the SIDM implementation in \textsc{OpenGadget3} to the public, with the hope that it proves useful to the SIDM community for future studies. In this spirit, we have tried to make it easy to implement additional features, such as differential cross-sections.
Other extensions and improvements of the SIDM module or related developments may eventually be released in the future.

\section*{Acknowledgments}
The authors thank the anonymous referees for helpful comments that
improved the paper.
They also thank all participants of the Darkium SIDM Journal Club for discussion. Moreover, they are very grateful to Felix Kahlhöfer for his long-term contribution in several publications developing and using the SIDM module of \textsc{OpenGadget3} and the supervision of MW's master thesis, which led to the implementation of full differential cross-sections described in Sect.~\ref{sec:differential_cross_section} and Appendix~\ref{sec:sampling_differential_cross_section}.
The authors also thank Sabarish Venkataramani and Lenard Kasselmann for their contribution and feedback on the SIDM module in \textsc{OpenGadget3} via several previous projects.
MSF thanks Geray Karademir for his support making the SIDM implementation in \textsc{OpenGadget3} public, and Hai-Bo Yu for his comment on cross-sections that should be included.
MSF gratefully acknowledges the support of the Alexander von Humboldt Foundation through a Feodor Lynen Research Fellowship.
KD and MSF acknowledge support by the COMPLEX project from the European Research Council (ERC) under the European Union’s Horizon 2020 research and innovation programme grant agreement ERC-2019-AdG 882679. MB and KSH acknowledge funding by the Deutsche Forschungsgemeinschaft (DFG) under Germany's Excellence Strategy -- EXC 2121 ``Quantum Universe" --  390833306. MB also acknowledges funding by and the DFG Research Group ``Relativistic Jets".
MG acknowledges support by the Excellence Cluster ORIGINS, which is funded by the Deutsche Forschungsgemeinschaft (DFG, German Research Foundation) under Germany’s Excellence Strategy - EXC-2094 - 390783311 as well as the DFG Collaborative Research Centre ``Neutrinos and Dark Matter in Astro- and Particle Physics'' (SFB 1258).
Preprint number: DESY-26-034.

Software:
NumPy \citep{NumPy},
Matplotlib \citep{Matplotlib}.

\section*{Data Availability}
The full \textsc{OpenGadget3} code is publicly available at \url{https://gitlab.lrz.de/AstroCodes/OpenGadget3} \citep{Dolag_2026}.
We note that the released version also contains the implementations underlying the works by \cite{Klemmer_2026} and by \cite{Schmidt_2026}.
Other data underlying this article will also be shared on reasonable request.
In addition, the data of \cite{Fischer_2025b}, that we reused in Sect.~\ref{sec:collapse} can be retrieved from our webpage: \url{https://www.darkium.org}.

\bibliographystyle{aasjournal}
\bibliography{bib}

\appendix
\numberwithin{figure}{section}
\numberwithin{table}{section}
\numberwithin{equation}{section}

\section{Sampling methods for differential cross-sections} \label{sec:sampling_differential_cross_section}

As mentioned in Sect.~\ref{sec:differential_cross_section}, there are two interpolation schemes for the purpose of simulating full differential cross-sections.
In this Appendix, we describe them in greater detail.

The problem of computing the scattering angle $\theta_\mathrm{cms}$ can be formulated in geometrical terms by using the cumulative distribution function (Eq.~\eqref{eq:cdf}) and thinking about the equation
\begin{equation}
\label{eq:cdf_x}
    \mathrm{CDF}(\theta_\mathrm{cms}(v,x)) = x\,,
\end{equation}
defining a surface in $(v, x, \theta_\mathrm{cms}) \in [0,\infty) \times [0,1] \times [\theta_\mathrm{min},\theta_\mathrm{max}]$.
For the purpose of interpolation, one must find a suitable triangulation of that surface. To restrict the number of available triangulations, we choose a regular grid in two of the three axes. The methods thus essentially differ in the choice of their base axes, which will also fundamentally change the interpolation routine. There are three combinations of base axes that one can choose from. However, a grid in $x$--$\theta$ is not useful because for any given combination of $x$ and $\theta$, the corresponding velocity $v$ may not exist.
In the following, we first consider the case of choosing the grid in the $x$--$v$ plane (Sect.~\ref{sec:Grid-method}) followed by the grid in the $\theta_\mathrm{cms}$--$v$ plane (Sect.~\ref{sec:ZSlice-method}).

\subsection{Grid method} \label{sec:Grid-method}

Here, we discuss the method where we quantise the values of $x$ and $v$, defining a regular grid in the $x$--$v$ plane. A good choice for $x$ are equally spaced points, $\mathbf{X}$, while for $v$ the points, $\mathbf{V}$, should be exponentially spaced due to the wide range of relative velocities encountered in a simulation. We define
\begin{equation}
\label{eq:velocity_spacing}
    f_v(i) = \exp\left( \ln(v_\mathrm{min}) + \frac{i \, \left[ \ln(v_\mathrm{max}) - \ln(v_\mathrm{min}) \right]}{N_v-2}\right)\,,
\end{equation}
with $i \in \mathbb{N}_0$ and set
\begin{align}
\label{eq:X_and_V_array}
    \mathbf{X} &= \left(\, 0, \frac{1}{N_x-1}, \frac{2}{N_x-1}, ... , 1 \right)\,,\nonumber\\
    \mathbf{V} &= \left(\, 0, f_v(0), f_v(1), ... , f_v(N_v - 3), f_v(N_v - 2) \right)\,,
\end{align}
with $N_x$ and $N_v$ being adjustable parameters that control the granularity of the grid. $\mathbf{V}$ starts at $0$ to catch all interactions below $v_\mathrm{min}$. A table of $\theta_\mathrm{cms}$ values is then populated by solving Eq.~\eqref{eq:cdf_x} for each point $(x,v) \in \mathbf{X} \times \mathbf{V}$.

Given the regular grid in the $x$--$v$ plane, $\theta_\mathrm{cms}(x,v)$ is easily computed for a specific pair $(x^*, v^*)$ with a standard bilinear interpolation scheme. Since the points in $x$ are equally spaced, the enveloping points needed for the interpolation can be directly computed for $x^*$, while the enveloping points for $v^*$ are found with a binary search algorithm. We note that given Eq.~\eqref{eq:velocity_spacing}, a direct computation for the enveloping points for $v^*$ is also possible.

\subsection{ZSlice method} \label{sec:ZSlice-method}

\begin{figure}
    \centering
    \includegraphics[width=\columnwidth]{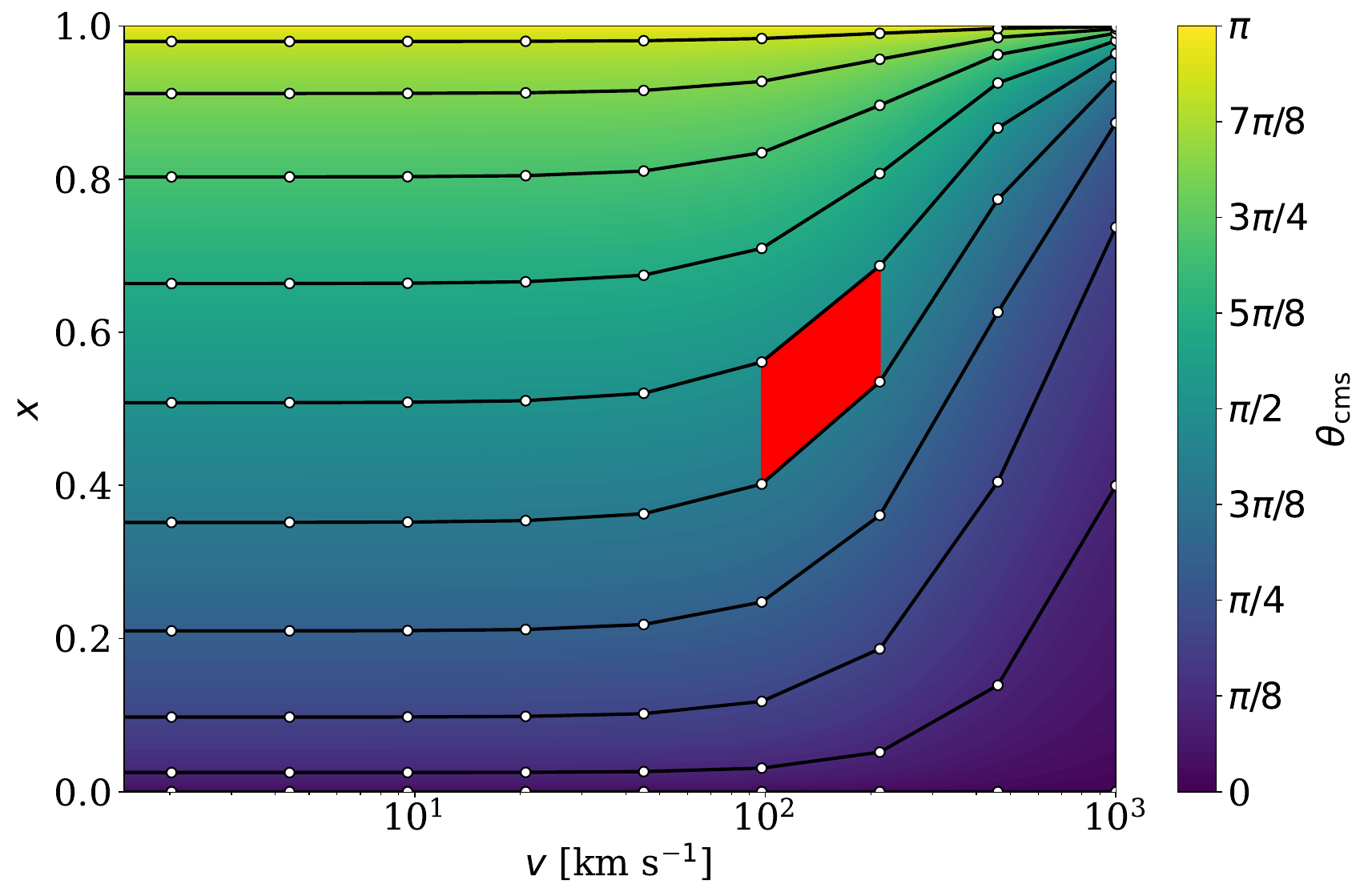}
    \caption{An illustration of the computed table using the ZSlice method. Shown is the computed Rutherford scattering angle with $w=200\,\mathrm{km} \, \mathrm{s}^{-1}$ and $N_x=N_h=100$. The connected points show the flow of equal scattering angle $\theta_\mathrm{cms}$. To keep the figure readable, only every tenth point has been drawn. One of the quadrilaterals encountered in interpolation is highlighted in red.}
    \label{fig:zslice_illustration}
\end{figure}

Here we consider a grid in the $\theta_\mathrm{cms}$--$v$ plane. The values for $v$ stay the same as for the grid method. The values, $\mathbf{\Theta}$, for $\theta_\mathrm{cms}$ are equally spaced from $\theta_\mathrm{min}$ to $\theta_\mathrm{max}$, similar to $\mathbf{X}$ given by Eq.~\eqref{eq:X_and_V_array}. Using $\Delta \theta = \theta_\mathrm{max} - \theta_\mathrm{min}$ we write
\begin{equation}
\label{eq:Theta_array}
    \mathbf{\Theta} = \left( \, \theta_\mathrm{min}, \theta_\mathrm{min} + \frac{\Delta \theta}{N_h-1}, \theta_\mathrm{min} + \frac{2\, \Delta \theta}{N_h-1}, ... , \theta_\mathrm{max} \,\right)\,,
\end{equation}
where $N_h$ controls the granularity of the points in $\theta_\mathrm{cms}$. We populate a table of $x$ values for each point $(\theta_\mathrm{cms}, v) \in \mathbf{\Theta} \times \mathbf{V}$, which is easily computed by numerically evaluating Eq.~\eqref{eq:cdf}. However, due to the fact that the value of $x$ gives the probability that the interacting particles scatter at an angle less than $\theta_\mathrm{cms}$ for any given $v$, the table generally no longer forms a regular grid in $x$ across the entire velocity range. Instead, they follow the contour of equal $\theta_\mathrm{cms}$, similar to equipotential lines. Figure~\ref{fig:zslice_illustration} illustrates the behaviour for the Rutherford cross-section (Eq.~\eqref{eq:rutherford_dcs}) with $w=200\,\mathrm{km} \, \mathrm{s}^{-1}$. This implies the need for a new interpolation scheme because the sampled points can easily lie outside the quadrilateral selected by bilinear interpolation.

\begin{figure}
    \centering
    \includegraphics[width=\columnwidth]{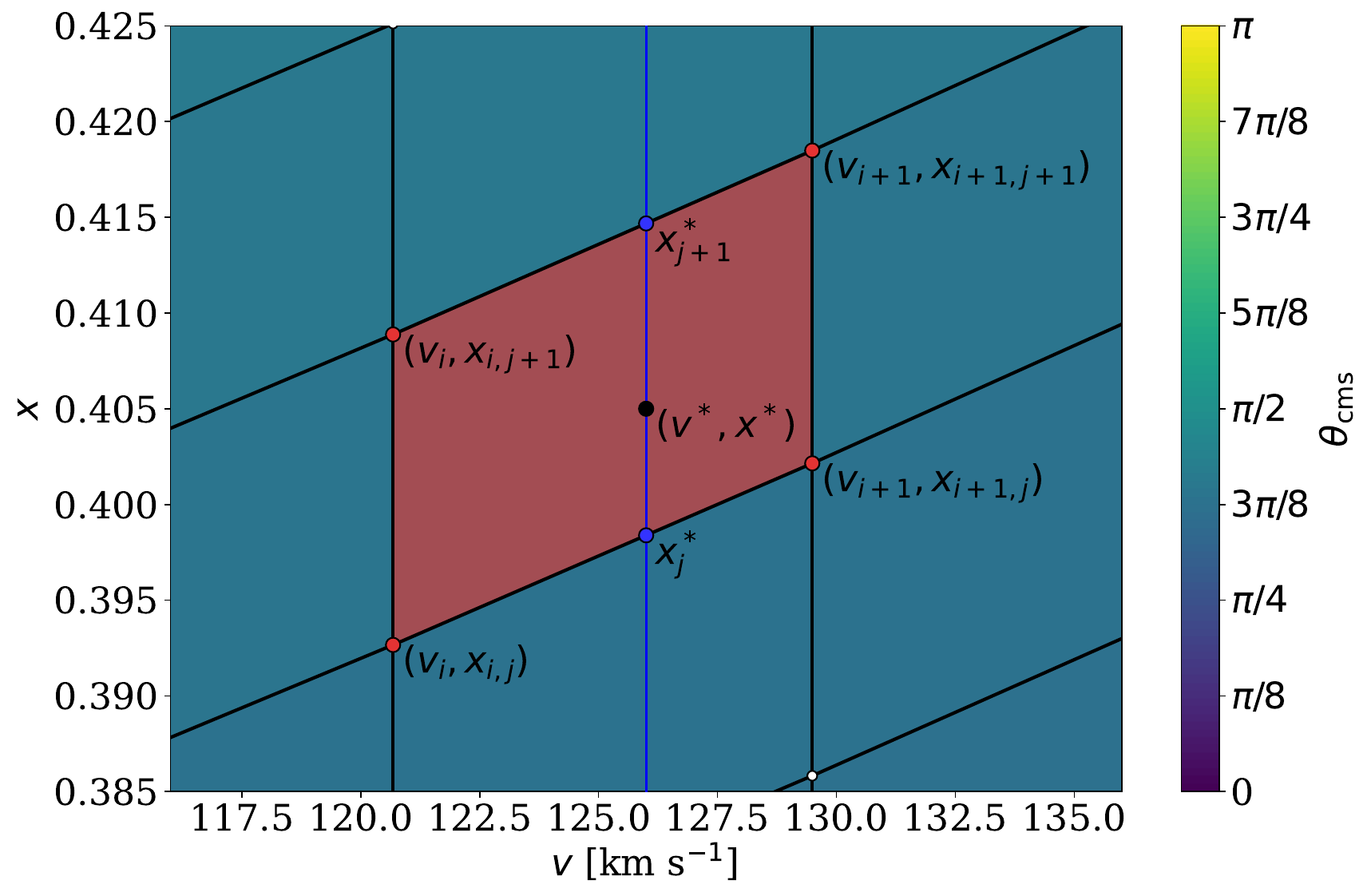}
    \caption{Example of the ZSlice interpolation scheme. One quadrilateral is highlighted in red. The interpolation for the velocity from it is indicated by the blue points. They are used for the final interpolation along the blue line to obtain the scattering angle. The resulting point is indicated by the black point.}
    \label{fig:zslice_quad}
\end{figure}

Consider the situation in Fig.~\ref{fig:zslice_quad}. To find the correct quadrilateral in which a given sample point $(x^*, v^*)$ is located, we first determine the nearest neighbours in velocity $v_i, v_{i+1}$ such that $v_i < v^* < v_{i+1}$. Further searching along the $x$-axis at $v_i$ we find the nearest left neighbours $x_{i, j}, x_{i, j+1}$ such that $x_{i,j} < x^* < x_{i,j+1}$. From this we can reconstruct the entire quadrilateral and find the right neighbours $(v_i, x_{i+1, j}), (v_{i+1}, x_{i+1, j+1})$, since these points have the same indices within the table as the left side. We compute the velocity interpolation factor
\begin{equation}
    \lambda = \frac{v^* - v_\mathrm{i}}{v_\mathrm{i+1} - v_\mathrm{i}}\,,
\end{equation}
and project the sample point along the $x$-axis to find $x^*_{j}, x^*_{j+1}$,
\begin{align}
    x^*_{j} &= x_{i,j} + \lambda \, (x_{i+1,j} - x_{i,j})\,, \nonumber\\
    x^*_{j+1} &= x_{i,j+1} + \lambda \, (x_{i+1,j+1} - x_{i,j+1})\,.
\end{align}
The sample point then lies within the quadrilateral, when
\begin{equation}
\label{eq:inside_quad}
    x^*_{j} \leq x^* \leq x^*_{j+1}\,.
\end{equation}
If that condition is fulfilled, $\theta_\mathrm{cms}$ follows by linearly interpolating between the upper and lower points,
\begin{align}
    \eta &= \frac{x^* - x^*_{j}}{x^*_{j+1} - x^*_{j}}\,, \nonumber\\
    \theta_\mathrm{cms} &= \theta_{\mathrm{cms},j} + \eta \left( \theta_{\mathrm{cms},j+1} - \theta_{\mathrm{cms},j} \right)\,,
\end{align}
with $\theta_{\mathrm{cms},j},\, \theta_{\mathrm{cms},j+1}$ being the enveloping scattering angles of $\mathbf{\Theta}$ (Eq.~\eqref{eq:Theta_array}).

Condition Eq.~\eqref{eq:inside_quad} may not hold in areas of steep changes in $\theta_\mathrm{cms}$. In this case, the quadrilateral reconstructed from the left points is no longer valid. We can repeat the same process but instead reconstruct the quadrilateral from the right side points first. Almost all points in parameter space are covered by either quadrilateral. However, for a small subset of points, the condition Eq.~\eqref{eq:inside_quad} does not hold for either reconstructed quadrilateral. In this case, they form an upper and lower boundary, and the true quadrilateral can be found by iterating through them and checking the condition.

\vspace{1ex}

\end{document}